\documentclass[usegraphicx,useAMS]{mn2e}

\newcommand{\lii}{Li\,{\footnotesize I}}

\newcommand{\kms}{\,km\,s$^{-1}$}

\newcommand{\be}{\begin{equation}}
\newcommand{\ee}{\end{equation}}
\newcommand{\bd}{\begin{displaymath}}
\newcommand{\ed}{\end{displaymath}}

\title[The Lithium depletion boundary of NGC 2547]
  {The Lithium Depletion Boundary in NGC 2547 as a test of
  pre-main-sequence evolutionary models\thanks{Based on observations
  collected with the VLT/UT2 Kueyen telescope (Paranal Observatory,
  ESO, Chile) using the FLAMES/GIRAFFE spectrograph (Observing run 072.D-0406A).}}
\author[R.D. Jeffries and J.M. Oliveira]
  {R.D.~Jeffries and J.M.~Oliveira\\
  Astrophysics Group, School of Chemistry and Physics, Keele University, Keele, 
      Staffordshire ST5 5BG, United Kingdom\\
}
\date{Submitted August 20 2004}

\pagerange{\pageref{firstpage}--\pageref{lastpage}} \pubyear{2004}

\def\LaTeX{L\kern-.36em\raise.3ex\hbox{a}\kern-.15em
    T\kern-.1667em\lower.7ex\hbox{E}\kern-.125emX}

\begin{document}

\label{firstpage}

\maketitle

\begin{abstract}
Intermediate resolution spectroscopy from the ESO Very Large Telescope is
analysed for 63 photometrically selected low-mass
(0.08-0.30\,$M_{\odot}$) candidates of the open cluster NGC 2547.  We
have confirmed membership for most of these stars using radial velocities, and
found that lithium remains undepleted for cluster stars with
$I>17.54\pm 0.14$ and $K_{s}>14.86\pm 0.12$. From these results, several
pre-main-sequence evolutionary models give almost model independent
ages of 34-36\,Myr, with a precision of 10 per cent. These ages are only
slightly larger than the ages of 25-35 ($\pm 5$)\,Myr obtained using the
same models to fit isochrones to higher mass stars descending towards
the zero age main sequence (ZAMS), both in empirically calibrated and
theoretical colour-magnitude diagrams. This agreement between age
determinations in different mass ranges is an excellent test of the
current generation of low-mass pre-main sequence stellar models and
lends confidence to ages determined with either method between
30 and 120\,Myr.

\end{abstract}


\section{Introduction}

Absolute ages in young open clusters and star forming regions are difficult
to determine and usually very model dependent, yet are crucial to solving
many stellar evolution problems -- from the lifetimes of protoplanetary
disks and progression of angular momentum loss, to calibrating cooling
models and the initial-final mass relation for white dwarfs.  Nuclear
turn-off ages from high mass stars may be uncertain by factors of two,
depending on the treatment of core convection, rotation and mass loss
(Chiosi, Bertelli \& Bressan 1992; Meynet et al. 1993; Meynet \& Maeder
1997, 2000). Ages from fitting isochrones to cool, low-mass stars, descending
towards the ZAMS may be equally inaccurate: they depend on detailed
modelling of stellar atmospheres, including molecular opacities, convection
and turbulence (see Baraffe et al. 1998, 2002) and
observational comparisons suffer from uncertainties in the conversion
from model parameters ($L_{\rm bol}$ and $T_{\rm eff}$) to magnitudes
and colours (e.g. Stauffer, Hartmann \& Barrado y Navascu\'es 1995).

The {\em Lithium Depletion Boundary} (LDB) is a new, important, and
possibly less model dependent method for determining the ages of young
clusters. The initial Li content of a low-mass star begins to burn in
$p,\alpha$ reactions, as stars contract along PMS tracks and core
temperatures reach $2.5\times10^{6}$\,K. Below 0.065$M_{\odot}$, Li is
never burned because the core does not become hot enough. At higher
masses, the time taken to reach Li-burning temperatures is a sensitive
function of mass and hence a {\em very sensitive function of
luminosity} (Bildsten et al. 1997; Ushomirsky et al. 1998).
Low-mass stars approaching the ZAMS are fully convective, so their entire
Li content is then rapidly exhausted (in less than a few Myr after Li
burning is initiated) and thus
in a cluster, the luminosity at which photospheric Li remains
undepleted can give a very precise age estimate and is most sensitive
for ages of 10-200\,Myr. 

The small experimental uncertainties in the LDB
method are due to the conversion from optical and near infrared magnitudes to
luminosities via cluster distances and empirical bolometric corrections,
and the difficulty in locating the LDB among a population of very
faint, low-mass stars (see Jeffries \& Naylor 2001). Systematic errors
are likely to be quite small. Remarkably, different evolutionary
models and even analytic calculations yield almost identical (to $\pm
10$\%) ages, despite very different treatments of atmospheres,
convection, opacities and equations of state (see for example Burke, Pinsonneault
\& Sills 2004).  It is this model independence which makes it vital to
measure the LDB in as many young clusters as possible. Such clusters
would define empirical isochrones that could be used to link together and
calibrate the uncertain physics in both high- and low-mass stellar
evolution models.

LDB ages have so far only been determined for the Pleiades ($125\pm8$\,Myr),
Alpha Per ($90\pm10$\,Myr) and IC\,2391 ($53\pm5$\,Myr) clusters
(Stauffer, Schultz \& Kirkpatrick 1998; Stauffer et al. 1999; Barrado y
Navascu\'es, Stauffer \& Patten 1999; Barrado y Navascu\'es, Stauffer
\& Jayawardhana 2004). These ages are older by factors of $\simeq1.5$ than
nuclear turn-off ages determined from high mass evolutionary models
with zero convective overshoot. The fractional discrepancy may become
smaller towards younger ages (and hence higher masses at the turn-off),
hinting that additional convective overshoot is mass-dependent.  

For IC\,2391, the LDB age is also about 15\,Myr older than the age
indicated by fitting isochrones to the low-mass ($0.3<M<1.2M_{\odot}$)
PMS cluster members.  This may also be true for the cluster NGC 2547,
which has a nuclear turn-off age of $55\pm25$\,Myr (Jeffries \& Tolley
1998), a low-mass isochronal age of 20-35\,Myr (Naylor et al. 2002) a
distance modulus of $8.10\pm 0.10$ (Naylor et al. 2002) and a
reddening, $E(B-V)=0.06\pm0.02$ (Clari\'a 1982). A previous attempt to
detect the LDB in NGC 2547 was made by Oliveira et al. (2003), but
resulted in lower limits to the cluster age. No Li-rich objects were
found for $I<17.2$ and by considering the average spectrum of several
cluster candidates there was some evidence that the LDB lay between
$17.8<I<18.3$. This suggested that the LDB age of the cluster was
certainly more than 30\,Myr and probably more like 38-46\,Myr and hence
older than the low-mass isochronal age. Oliveira et al. suggested that
this discrepancy might indicate shortcomings in the commonly used
low-mass evolutionary models.

In this paper we present optical spectroscopy from the ESO Very Large
Telescope (VLT) that unambiguously identifies a substantial population
($\sim 50$)
of very low-mass stars in NGC 2547 and now 
precisely determines the location of the LDB.
In section 2 we discuss the selection of targets, the
fibre spectroscopy performed at the VLT and the data reduction
process. In section 3 we report measurements of relative radial
velocities and the \lii~6708\AA\ feature in the cluster candidates, and
our assessment of cluster membership. Section 4 deals with the location
of the LDB and the age of NGC 2547. The results are discussed in
section 5 and conclusions drawn in section 6.

\section{VLT Fibre Spectroscopy}

\subsection{Target selection}

The targets for VLT fibre spectroscopy were chosen from the $RIZ$
survey of Jeffries et al. (2004). This catalogue contains precise
astrometric and photometric information for stellar objects in 0.855
square degrees around NGC~2547 and is substantially complete to
$I=19.5$.  Fifty four targets were chosen from the lists of candidate
cluster members with $16.4<I<19.1$ which were in turn selected from
their photometric colours and proximity to empirically calibrated
30\,Myr isochrones generated from the models of D'Antona \& Mazzitelli
(1997) and Baraffe et al. (2002) (see Jeffries et al. 2004 for
details). Using an intrinsic distance modulus of 8.1 and extinction,
$A_{I}=0.112$ (Rieke \& Lebofsky 1985), cluster members in this
magnitude range will have approximate masses of 0.08-0.30\,$M_{\odot}$
according to the models of Baraffe et al. (2002).

These targets are only a small, representative subset of a larger
 population of NGC 2547 cluster members in this magnitude range, that
 is spread over more than a square degree.  A further nine objects were
 observed that were slightly more distant from the cluster isochrones and not
 classed as members by Jeffries et al., but which were also contained
 within the 25 arcminute diameter VLT field of view.  The precision of
 the photometry is about $\pm 0.04$ in $I$ and $\pm 0.07$ in $R-I$ for
 the faintest object in the sample (with $I=19.05$), but is more like
 $\pm (0.01-0.02)$ in colour and magnitude for the majority.

\subsection{Observations}

The targets were observed on 19 and 20 January 2004 
using the FLAMES instrument mounted on the
VLT-Kueyen (UT2) 8.2-m telescope. The GIRAFFE spectrograph and MEDUSA
fibre system were used in combination with the 600 lines\,mm$^{-1}$
grating and an order sorting filter to give spectra with a resolving power
of about 8000 over the wavelength range 6438\AA\,$< \lambda < 7184$\AA.
The MEDUSA system has 130 fibres, each with a projected diameter of 1.2
arcseconds on the sky, that can be positioned within a circular area of
diameter 25 arcminutes.

The targets were observed in 8 separate observing blocks (OBs), each of which
yielded 2589\,s of observing time on sky and all of which were centred
at \mbox{RA\,$=$\,08h\,10m\,06.6s,} \mbox{Dec\,$=-$49d\,15m\,43s.} A number of fibre
configurations were used, such that most objects (and all with
$I>18.5$) were observed in all 8 OBs, a few targets were observed in
5 or 6 OBs and one target (number 35 - see
Table~\ref{main}) was observed in only 2 OBs.  This was done to maximise
the number of targets observed at the required level of precision. Six
of the OBs were obtained on 19 January 2004 between 03:14\,UT and
08:43\,UT. 
The other two OBs were obtained on 20 January 2004 between
03:09\,UT and 07:28\,UT. About 50 fibres which could not be allocated to
useful targets were placed on blank sky regions in each
configuration. The seeing measured at the telescope was 0.5-0.9
arcseconds and the moon was below the horizon during the observations.
Flat-field, bias and Thorium-Argon arc lamp calibrations were performed
at the start of each observing night.

\subsection{Data Reduction}

The calibrations and observational data were reduced using version 1.10
of the GIRAFFE pipeline software (Blecha et al. 2003). Master bias and
flat-field calibrations were produced for each night along with fits
tables containing the locations of the fibre spectra (from the master
flat-field) on the 2K$\times4$K EEV CCD detector and wavelength
calibration solutions for all the fibres taken from the arc lamp
spectra. Each target exposure was bias-subtracted, and scattered light
was removed by modelling the remaining signal between the individual
spectra. Target and blank sky spectra were extracted using an optimal
algorithm. Pixel-to-pixel and fibre-to-fibre sensitivity variations
were corrected for by dividing by a normalised set of spectra extracted
from the master flat-field using the same apertures. Note that no
attempt is made to flux calibrate the spectra in an absolute sense, but
that most of the instrumental response is removed by dividing by the
flat-field continuum. The corrected
spectra were then rebinned to a wavelength scale of 0.2\AA\ per pixel,
slightly larger than the average wavelength coverage of a single
detector pixel.

Sky subtraction was achieved by taking the median of groups of five
blank sky spectra and averaging these medians to form a master sky
spectrum which could be subtracted from each target spectrum. The
standard deviation among the continuum levels in the individual sky
spectra was about 7 per cent (1-sigma). Part of this arises from
uncertainties in the fibre-to-fibre sensitivity corrections, resulting
largely from non-uniform illumination of the fibres by the flat-field
lamp (L. Pasquini, ESO, private communication). There may also be a
contribution from scattered continuum radiation from a bright, {\it
spatially varying}, H\,{\sc ii} region which coincides with the NGC
2547 line of sight and is clearly visible in $R$-band images. The
H\,{\sc ii} region is even more evident in the bright night-sky
emission lines, where variations of 20-30 per cent are found even between
sky fibres situated only a few arcminutes apart.

Uncertainties in the subtracted sky continuum are of no concern for
this paper as far as measuring the continuum level close to the
\lii\,6708\AA\ feature.  The sky continuum is just equal to the signal
from the targets at $I\simeq 19$ and hence for an unresolved line of width $\simeq
0.8$\AA\ there may be an additional equivalent width error of about
0.05\AA, which is smaller than the statistical errors in the equivalent
width measurements even for the faintest targets (see section 3.3).
However, the strong sky lines are up to 200 times stronger than
the sky continuum and so sky-subtraction in the immediate vicinity of
these wavelengths (H$\alpha$, S\,{\sc ii} 6717\AA, 6731\AA\ etc.) was
more problematic.

\begin{figure*}
\includegraphics[width=160mm]{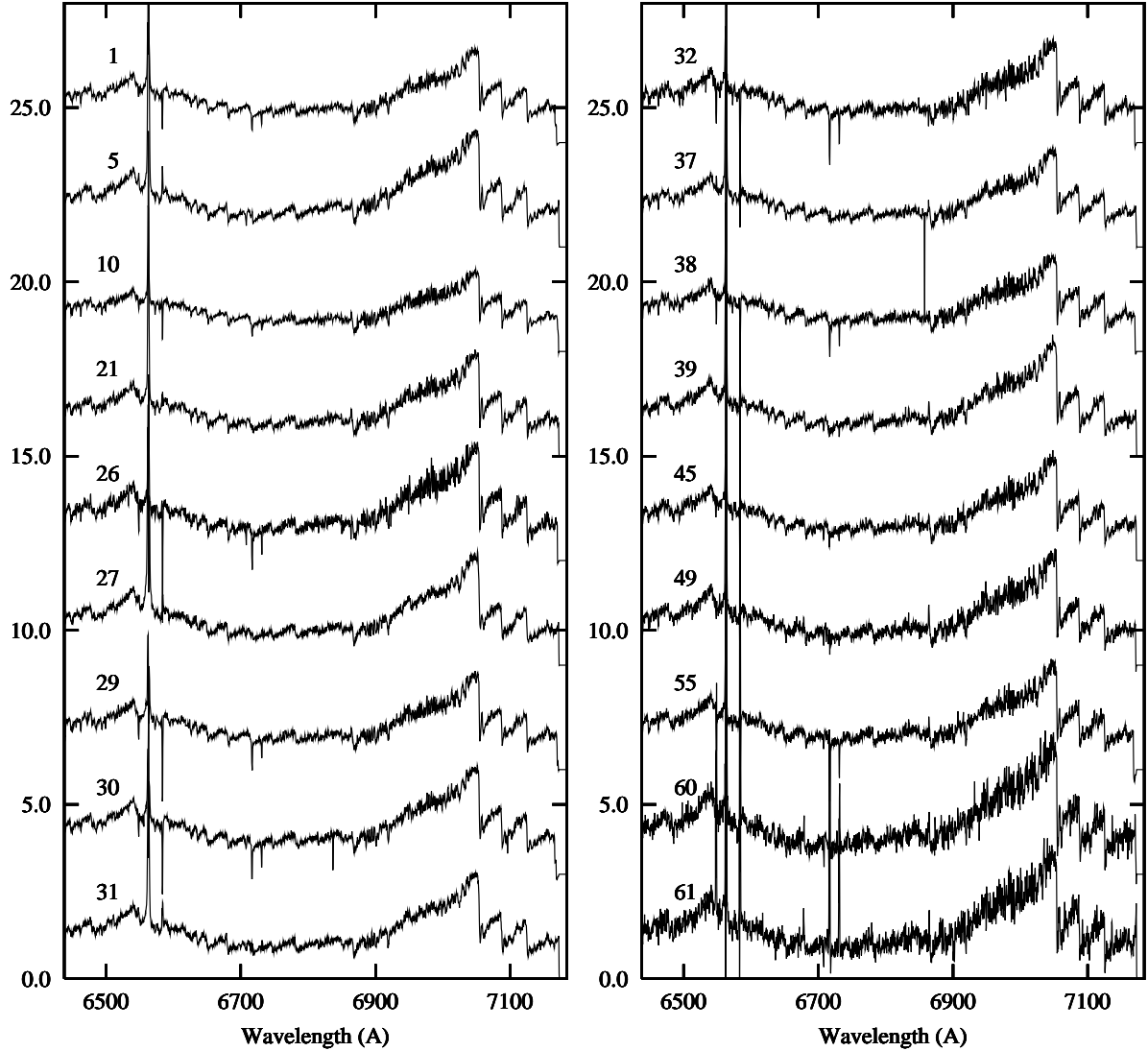}
 \caption{A representative sample of our reduced spectra covering the
 full range of colours and magnitudes. The spectra are normalised to
 unity at 6705\AA\ and offset. The labels
 refer to the identifiers in Table~\ref{main}.}
    \label{specmontage1}
\end{figure*}

Multiple spectra for targets were combined using an iterative technique
that rejected pixels deviating by more than 5-sigma from from the
average of pixels at the same wavelength in all the normalised spectra
of that target.  Three separate combinations were produced: all of the
spectra; all spectra from 19 January 2004; and all spectra from 20
January 2004. Where only two spectra of an object existed in the
combination, these spectra were simply summed. The theoretical levels
of signal-to-noise ratio (SNR) achieved in the averaged spectra from
all the target exposures, based on measurements of the CCD gain and
readout noise and propagating errors in the bias subtraction,
flat-fielding, scattered light subtraction and sky subtraction, ranged
from 6 to more than 50 per 0.2\AA\ pixel in the faintest and brightest
targets respectively.  A sample of our spectra covering the full range
of wavelengths, colours and magnitudes is shown in
Fig.~\ref{specmontage1}.

\begin{figure*}
\includegraphics[width=160mm]{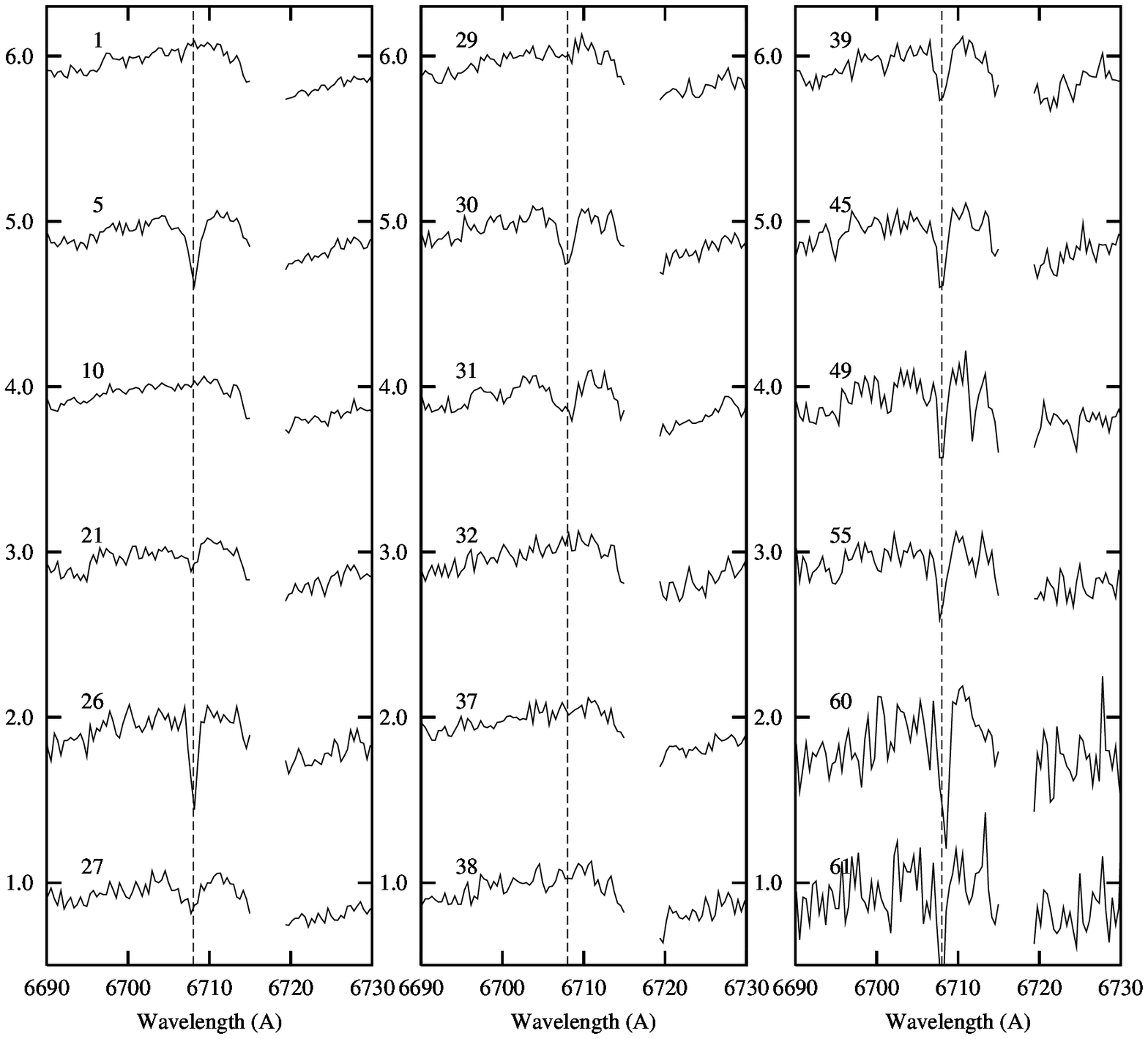}
 \caption{Representative spectra in the vicinity of the Li\,{\sc i}
 6708\AA\ spectral feature. The spectra are labelled according to the
 identifiers in Table~\ref{main} and plotted in 0.4\AA\ bins. 
 The gaps in the spectra are the
 location of the prominent S\,{\sc ii} line found in the sky spectra,
 which could not be accurately subtracted. The spectra are normalised
 to unity around 6705\AA\ and offset. The vertical dashed line
 indicates the expected position of the Li\,{\sc i} 6708\AA\ feature
 for members of NGC 2547.}
    \label{limontage}
\end{figure*}

\section{Analysis}
\subsection{Spectral Indices}
\label{specindex}

To check that all the observed targets had spectral types appropriate
for their colours, we calculated the TiO\,(7140\AA) and CaH\,(6975\AA)
narrow band spectral indices (see Oliveira et al. 2003). These indices are
temperature sensitive and we expect a smooth relationship between their
values and the $R-I$ colour index of NGC 2547 cluster members.

None of our targets are found to deviate significantly from the trend
defined by the bulk of the data. However, this is not unexpected. Any
contamination in our sample is certain to be dominated by foreground
field M-dwarfs with almost identical narrow band indices at the same
colour (Jeffries et al. 2004). The CaH index does have some sensitivity
to gravity but this is unlikely to distinguish between cool
pre-main-sequence objects at 30-50\,Myr and main-sequence
dwarfs. Indeed, targets which are deemed cluster non-members by virtue
of their relative radial velocities (see section~\ref{rv}) are
indistinguishable from cluster members on the basis of these narrow
band indices. Table~1 lists spectral types for our targets, estimated
by comparing their TiO\,(7140\AA) narrow band indices with the those
measured for stars with known spectral type (see Oliveira et
al. 2003). We estimate these spectral types are good to $\pm$ half a
subclass.

\subsection{Radial velocities}
\label{rv}

Although we were unable to obtain observations of radial velocity
standard stars during our observations, the {\it relative} radial
velocities (RVs) of our targets proved a useful tool to
discriminate cluster members from non-members and also identify short
period (of order 10 days or less) binary systems in those targets which
show other indications of membership (i.e. a detection of Li -- see
section~\ref{li}). 
Members of the cluster (bar short period binaries) should have RVs with
a dispersion of order 1\,\kms\ (Jeffries, Totten \& James 2000), 
so given that we expect the bulk of our
targets to be cluster members,
selection in a narrow range of relative RVs should exclude the vast
majority of non-members.

Radial velocities were calculated by cross-correlating the region
$6950< \lambda <7150$\AA\ which includes prominent molecular band heads and only a
few very weak nebular emission lines and telluric features.
We chose one of the targets (number 30 -- see Table~\ref{main}) 
with a good SNR and an intermediate colour
and spectral type to act as the RV template for all the other objects.
All the spectra from each night were cross-correlated with the spectrum
of object 30 obtained on 19 January 2004. Hence most objects have two relative RV
measurements. 

The average relative RV for the cluster
members and the precision of the RV measurements were estimated by
considering the subsample of targets that were Li-rich and hence almost
certain cluster members (see section~\ref{membership}), but
excluding three objects which had clearly discrepant or varying RVs
from night to night. We found an average RV of $(-0.9\pm 0.3)$\kms\
with a standard deviation of 1.7\kms\ on 19 January 2004 and an average
RV of ($+0.0\pm 0.6)$\kms with a standard deviation of 2.6\kms\ on 20
January 2004. As the combined data from 19 January had a higher SNR,
the smaller standard deviation is expected. 

Cluster members were required to have RVs
within $n\sigma$ of the average RV of the Li-rich objects on 19 January
2004 and a consistent RV measured on 20 January 2004. The
value of $n$ was adjusted depending on the empirical SNR (defined in
the next section) of the
spectrum in question. For SNR\,$>20$, we used $n=2$, for $10<$\,SNR\,$<20$, $n=3$ and
for SNR\,$<10$, $n=4$. Before comparing the RVs from 19 and
20 January 2004, the mean discrepancy of 0.9\kms\ was subtracted from
the data taken on 20 January. The relative RVs for the targets on each
night are listed in Table~\ref{main}.

This process yielded 52 objects deemed to be cluster members, 9
objects deemed to be RV non-members and a further 2 objects where the
RV appeared to be variable between data from the two nights. For some
targets the cross-correlation function was markedly broadened, perhaps
as a result of rapid rotation (see below) or binarity. These RV
measurements are indicated in Table~\ref{main}.

\subsection{Lithium}\label{li}

The main aim of our experiment was to search for Li in the form of the
\lii\,6708\AA\ resonance doublet. This feature should be strong in cool
stars with undepleted Li -- with an equivalent width (EW) of 0.5\AA\ to
0.6\AA\ according to the curves of growth presented by Zapatero-Osorio
et al. (2002). Figure~\ref{limontage} shows a selection of our spectra
in this wavelength region, where data combined from both nights are shown.

The EW of the 6708\AA\ feature was estimated with respect to a
pseudo-continuum defined by a linear fit to the data in the wavelength
ranges 6700 to 6705\AA\ and 6710 to 6713\AA. The upper limit to this
latter continuum region was chosen to coincide with a
downturn in the spectra due to a molecular bandhead, but also to avoid
regions within a few angstroms of the S\,{\sc ii} 6717\AA\ sky line,
which was poorly subtracted in many of our spectra and contained very
noisy data in any case. The residuals to this linear
fit gave an empirical estimate of the SNR which was used in turn to
estimate either an uncertainty in the Li EW or a 2-sigma upper limit
where no line was seen. The empirical SNR is bound to underestimate the
true SNR of the spectra because there are molecular features in the
pseudo-continuum that become stronger at cooler
temperatures. Hence our EW uncertainty estimates and upper limits 
should be conservative and our EWs for the coolest stars may be
systematically underestimated.

The EWs were estimated by fitting and integrating a Gaussian absorption
feature beneath the linear continuum. Uncertainties were estimated
according to the approximate formula
\begin{equation}
\label{snrequation}
\Delta EW = \frac{\sqrt{{\rm FWHM}\times p}}{SNR}\, ,
\end{equation}
where FWHM and $p$ are the FWHM of the Gaussian and the size of the
pixels in \AA. The Li EWs or 2-sigma upper limits are listed in Table~\ref{main} along
with the empirical SNR estimate. For target 63, which showed a
large RV shift between the two nights, the quoted results are
from the first night.

Most of our targets have FWHM in the range 0.7 to 1.0\AA\ as
expected from the resolving power of the spectrograph. A few objects
showed Li lines which were much broader than this, with FWHM of
1.5-2.2\AA. This is likely due to rapid rotation either in single
objects simply because they are young (a phenomenon certainly seen in
higher mass objects in the cluster -- Jeffries et al. 2000)
or because they are part of close, tidally locked binary systems.
These broadened objects are indicated in Table~\ref{main}. The data are
of insufficient quality to estimate projected rotational velocities,
but they must be of order 50\kms\ to produce the broadening observed.

\begin{table*}
\tiny
\caption{Targets observed in this paper. The columns list identifiers
  used in this paper, identifiers from the catalogue of Jeffries et
  al. (2004) a ``N'' following the name indicates it was {\em not}
  considered a photometric member in that paper, positions, $I$
  magnitudes, $R-I$ and $I-Z$ colours from Jeffries et al. (2004),
  $JHK_{s}$ magnitudes from 2MASS (Cutri et al. 2003), estimated
  spectral type (see section 3.1), relative radial
  velocity (nights 1 and 2, a ``b'' indicates a broadened
  cross-correlation function -- see section~\ref{rv}), whether the
  radial velocity appears variable (V), the empirical SNR of the
  spectra around the Li~6708\AA\ feature, the measured equivalent
  widths for this feature or a 2-sigma upper limit (a ``b'' indicates
  the line is broadened -- see section~\ref{li})
  and our membership assessment status for the
  object -- (1) radial velocity member with no Li, (2) radial velocity
  member with Li, (3) radial velocity non-member with Li, (4) radial
  velocity non-member with no Li.  Photometric uncertainties are 0.01
  to 0.04 mag in $I$ and $I-Z$, 0.01 to 0.06 mag in $R-I$ and 0.04 to
  0.20 mag in $JHK_{s}$ (from brightest to faintest).}
\label{main}
\begin{tabular}{@{}rr@{\hspace*{1mm}}r@{\hspace*{1mm}}r@{\hspace*{2mm}}cccccccccrr@{\hspace*{0mm}}c@{\hspace*{0mm}}cl@{\hspace{3mm}}c}

\hline
ID& \multicolumn{2}{r}{J04 ID}& 
&RA&DEC&$I$&$R-I$&$I-Z$&$J$&$H$&$K_{s}$&SpT&RV$_1$&RV$_2$&
\multicolumn{1}{c}{Var?}&SNR&Li EW&Status\\
  &&        &&\multicolumn{2}{c}{(J2000)}&&&&&&&&\multicolumn{2}{c}{(km\,s$^{-1}$)}&&&(\AA)& \\
\hline
& & & & & & & & && & & & & &&&&\\
 1 &  17&  629 & &  8 09 17.721  & $-$49 08 34.70 &  16.508 &  1.751 &  0.496 &14.714  &14.212&  13.846 &M4.5&  $-$0.1 & +0.1 & & \llap{3}0 & \llap{$<$\,}0.03 &1\\  
 2 &   9&  467 & &  8 11 05.423  & $-$49 20 54.56 &  16.524 &  1.715 &  0.496 &14.767  &14.164&  13.856 &M4.5&  $-$2.0 & $-$3.2 & & \llap{3}4 & \llap{$<$\,}0.03 &1\\
 3 &  13&  605 & &  8 10 50.499  & $-$49 16 24.01 &  16.565 &  1.618 &  0.432 &14.907  &14.315&  14.019 &M4.5& \llap{+3}8.4 &\llap{+3}8.1 & & \llap{4}1 & \llap{$<$\,}0.02 &4\\
 4 &  13&  586 & &  8 09 29.032  & $-$49 14 33.13 &  16.644 &  1.627 &  0.460 &15.069  &14.427&  14.202 &M4.0&  $-$2.2 & $-$1.0 & & \llap{5}0 & \llap{$<$\,}0.02 &1\\
 5 &   8&  515&N &  8 10 11.757  & $-$49 21 01.94 &  16.648 &  1.955 &  0.609 &14.592  &14.012&  13.690 &M5.5&   +6.3 & +6.5 & & \llap{2}7 & $0.61\pm 0.03$\rlap{\,b} &3\\
& & & & & & & & && & & & & &&&&\\ 									       									      									      
 6 &  13&  712 & &  8 09 32.461  & $-$49 11 13.09 &  16.697 &  1.687 &  0.499 &14.894  &14.421&  14.057 &M4.5&  $-$1.3 & $-$1.2 & & \llap{3}0 & \llap{$<$\,}0.03 &1\\
 7 &  13&  736 & &  8 09 35.487  & $-$49 13 03.54 &  16.742 &  1.739 &  0.518 &15.008  &14.445&  14.071 &M4.5&  $-$2.8 & $-$1.1 & & \llap{2}0 & \llap{$<$\,}0.04 &1\\
 8 &  12&  621 & &  8 11 02.692  & $-$49 11 08.69 &  16.753 &  1.693 &  0.446 &15.084  &14.422&  14.076 &M4.5&  $-$2.0 & $-$1.5 & & \llap{3}1 & \llap{$<$\,}0.03 &1\\
 9 &   8&  554 & &  8 10 05.623  & $-$49 26 25.49 &  16.763 &  1.664 &  0.461 &15.010  &14.360&  14.133 &M4.5&  $-$2.8 & $-$2.0 & & \llap{3}0 & \llap{$<$\,}0.03 &1\\
10 &  12&  626 & &  8 10 53.561  & $-$49 11 43.20 &  16.816 &  1.607 &  0.448 &15.217  &14.703&  14.238 &M4.0&  $-$1.0 & $-$0.6 & & \llap{4}1 & \llap{$<$\,}0.02 &1\\
& & & & & & & & & && & & & &&&&\\ 									       									      									      
11 &  18&  729 & &  8 10 12.134  & $-$49 04 31.80 &  16.864 &  1.698 &  0.465 &15.234  &14.549&  14.205 &M4.5&  $-$1.0 & $-$2.1 & & \llap{2}7 & \llap{$<$\,}0.03 &1\\
12 &  13&  715 & &  8 10 25.353  & $-$49 11 18.68 &  16.918 &  1.834 &  0.558 &14.982  &14.433&  14.131 &M5.0&  $-$1.7 & $-$1.3 & & \llap{2}4 & \llap{$<$\,}0.04 &1\\
13 &  18&  738 & &  8 10 02.437  & $-$49 05 13.40 &  16.935 &  1.648 &  0.449 &15.269  &14.658&  14.351 &M4.5&  $-$1.9 & $-$0.9 & & \llap{2}8 & \llap{$<$\,}0.03 &1\\
14 &  14&  884 & &  8 09 11.922  & $-$49 14 43.11 &  16.946 &  1.745 &  0.481 &15.177  &14.629&  14.422 &M4.5&  $-$0.8 & $-$2.6 & & \llap{2}1 & \llap{$<$\,}0.04 &1\\
15 &  13&  676 & &  8 10 49.791  & $-$49 08 20.02 &  16.988 &  1.701 &  0.491 &15.245  &14.714&  14.414 &M4.5&  $-$1.8 & $-$1.5 & & \llap{2}4 & \llap{$<$\,}0.04 &1\\
& & & & & & & & && & & & & &&&&\\ 									       									      									      
16 &   7&  681 & &  8 09 09.250  & $-$49 20 42.77 &  16.999 &  1.644 &  0.469 &15.358  &14.761&  14.404 &M4.5&  $-$2.7 & $-$0.5 & & \llap{1}3 & \llap{$<$\,}0.07 &1\\
17 &  12&  863 &N&  8 10 46.949  & $-$49 15 21.64 &  17.038 &  1.607 &  0.486 &15.290  &14.645&  14.349 &M4.5&  $-$2.5 & $-$1.8 & & \llap{2}1 & \llap{$<$\,}0.04 &1\\
18 &  12&  738 & &  8 10 52.997  & $-$49 07 16.53 &  17.061 &  1.802 &  0.548 &15.208  &14.605&  14.174 &M5.5&  +0.9 & +1.8 & & \llap{1}7 & \llap{$<$\,}0.06 &1\\
19 &  14&  918 & &  8 09 13.184  & $-$49 17 38.79 &  17.077 &  1.711 &  0.499 &15.326  &14.737&  14.506 &M4.5&  $-$1.8 & +0.1 & & \llap{2}4 & \llap{$<$\,}0.04 &1\\
20 &  13&  892 & &  8 10 14.521  & $-$49 10 23.64 &  17.125 &  1.732 &  0.515 &15.299  &14.681&  14.504 &M4.5&  $-$0.3 & $-$2.0 & & \llap{2}2 &\llap{$<$\,}0.04 &1\\
& & & & & & & & && & & & & &&&&\\ 									       									      									      
21 &   8&  604 & &  8 10 21.472  & $-$49 17 48.61 &  17.149 &  1.799 &  0.536 &15.300  &14.670&  14.458 &M5.0&  $-$2.1 & $-$5.7 & & \llap{2}5 & $0.16\pm 0.03$\rlap{\,b} &1\\
22 &  13&  931 & &  8 09 51.086  & $-$49 12 18.48 &  17.162 &  1.751 &  0.504 &15.332  &14.697&  14.496 &M4.5&  $-$0.9 & $-$2.7 & & \llap{1}9 & \llap{$<$\,}0.05 &1\\
23 &   8&  679 &N&  8 10 17.675  & $-$49 23 29.74 &  17.165 &  1.626 &  0.449 &15.476  &14.802&  14.622 &M4.0&   +8.6 & +7.9 & & \llap{1}8 & \llap{$<$\,}0.05 &4\\
24 &   7&  727 &N&  8 09 24.619  & $-$49 24 17.27 &  17.179 &  1.631 &  0.422 &15.539  &14.989&  14.699 &M4.0&  \llap{+4}8.1 & \llap{+4}6.7 & & \llap{2}8 & \llap{$<$\,}0.03&4\\
25 &  13&  922 & &  8 10 04.600  & $-$49 11 39.09 &  17.210 &  1.749 &  0.506 &15.308  &14.784&  14.601 &M4.5&  $-$2.7 & $-$1.7 & & \llap{2}2 & \llap{$<$\,}0.04 &1\\
& & & & & & & & && & & & & &&&&\\ 									       									      									      
26 &  12&  752 & &  8 10 58.886  & $-$49 08 05.92 &  17.225 &  1.897 &  0.564 &15.364  &14.721&  14.331 &M5.5&  +2.9 &      & & \llap{1}7 & $0.59\pm 0.03$  &2\\
27 &   8&  655 & &  8 10 46.890  & $-$49 21 34.10 &  17.226 &  1.867 &  0.593 &15.184  &14.649&  14.291 &M5.5&  $-$1.5 & $-$0.3 & & \llap{1}9 & $0.36\pm 0.04$\rlap{\,b} &2\\
28 &  13& 1148 & &  8 09 56.050  & $-$49 11 16.46 &  17.233 &  1.731 &  0.515 &15.463  &15.029&  14.556 &M4.5&  $-$1.1 & $-$2.3 & & \llap{1}8 & \llap{$<$\,}0.05 &1\\
29 &  12&  810 & &  8 11 07.003  & $-$49 12 23.83 &  17.241 &  1.724 &  0.481 &15.485  &14.930&  14.482 &M4.5&  $-$1.6 & $-$1.0 & & \llap{2}0 & \llap{$<$\,}0.04 &1\\
30 &  13&  906 & &  8 10 03.099  & $-$49 11 02.76 &  17.252 &  1.948 &  0.559 &15.273  &14.675&  14.353 &M5.0&  +0.0 & +1.3 & & \llap{1}9 & $0.47\pm 0.04$\rlap{\,b} &2\\
& & & & & & & & && & & & & &&&&\\ 									       									      									      
31 &   7&  694 & &  8 09 50.218  & $-$49 21 15.89 &  17.291 &  1.819 &  0.556 &15.486  &14.782&  14.540 &M5.0&  $-$0.1 & $-$1.5 & & \llap{1}8 & $0.49\pm 0.04$\rlap{\,b} &2\\
32 &  12&  774 & &  8 11 11.865  & $-$49 10 16.63 &  17.304 &  1.800 &  0.468 &15.507  &15.015&  14.721 &M5.0&  $-$0.5 & $-$0.9 & & \llap{1}9 & \llap{$<$\,}0.06 &1\\
33 &   9&  760 & &  8 10 36.428  & $-$49 17 01.34 &  17.309 &  1.725 &  0.493 &        &      &         &M4.5&  +8.5 & +8.8 & & \llap{2}2 & \llap{$<$\,}0.04 &4\\
34 &  14& 1143 & &  8 09 23.534  & $-$49 16 27.44 &  17.318 &  1.674 &  0.541 &15.537  &14.923&  14.675 &M4.5&  $-$1.7 &      & & \llap{1}1 & \llap{$<$\,}0.08 &1\\
35 &  12&  792 & &  8 11 11.678  & $-$49 11 12.35 &  17.331 &  1.766 &  0.497 &15.571  &15.053&  14.203 &M4.5&  $-$1.7 &      & & \llap{1}0 & \llap{$<$\,}0.08 &1\\
& & & & & & & & && & & & & &&&&\\ 									       									      									      
36 &  12& 1127 & &  8 11 05.846  & $-$49 16 25.31 &  17.398 &  1.812 &  0.496 &15.481  &14.898&  14.575 &M5.0&  $-$2.2 & $-$4.5 & & \llap{1}7 & \llap{$<$\,}0.06 &1\\
37 &  14& 1530 & &  8 09 44.267  & $-$49 18 43.97 &  17.398 &  1.734 &  0.511 &15.509  &15.126&  14.879 &M5.0&  $-$0.3 & $-$2.4 & & \llap{2}2 & \llap{$<$\,}0.04  &1\\
38 &  12& 1000 & &  8 11 03.987  & $-$49 10 02.00 &  17.489 &  1.719 &  0.472 &15.826  &15.232&  14.756 &M4.5&  $-$1.8 & $-$0.9 & & \llap{1}8 & \llap{$<$\,}0.06 &1\\
39 &   8&  770 & &  8 10 12.072  & $-$49 17 41.61 &  17.576 &  1.797 &  0.557 &15.754  &15.008&  14.654 &M5.0&  $-$0.8 & $-$0.9 & & \llap{1}6 & $0.44\pm 0.04$\rlap{\,b} &2\\
40 &   7&  909 &N&  8 09 12.400  & $-$49 22 28.02 &  17.653 &  1.684 &  0.440 &15.973  &15.356&  15.087 &M4.5& \llap{+1}1.2 & +5.6 & & \llap{2}1 & \llap{$<$\,}0.04 &4\\
& & & & & & & & && & & & & &&&&\\ 									       									      									      
41 &  18& 1436 & &  8 10 16.488  & $-$49 03 36.78 &  17.692 &  1.854 &  0.584 &15.636  &15.125&  14.851 &M6.0&  $-$3.0 & $-$0.2 & & \llap{1}0 & $0.41\pm 0.04$  &2\\
42 &  12& 1202 & &  8 10 46.139  & $-$49 06 28.53 &  17.721 &  1.784 &  0.532 &16.143  &15.450&  15.030 &M4.5&  $-$1.0 & $-$1.5 & &  5 & $0.76\pm 0.10$  &2\\
43 &   9& 1011 & &  8 11 09.442  & $-$49 18 45.47 &  17.757 &  1.752 &  0.495 &16.227  &15.472&  14.861 &M4.5& \llap{+1}6.7\rlap{\,b}& \llap{+1}6.3 & & \llap{1}4 & \llap{$<$\,}0.06 &4\\
44 &  14& 1473 & &  8 09 29.481  & $-$49 16 40.12 &  17.803 &  1.763 &  0.535 &15.947  &15.308&  14.991 &M5.0&  $-$3.9\rlap{\,b}&  0.0\rlap{\,b}& & \llap{1}9 & $0.44\pm 0.10$\rlap{\,b}&2\\
45 &   9& 1039 & &  8 10 57.646  & $-$49 19 58.28 &  17.844 &  1.804 &  0.512 &16.068  &15.197&  15.176 &M5.0&  $-$0.9 & +0.4 & & \llap{1}4 & $0.50\pm 0.03$  &2\\
 & & & & & & & & 7& & & & &&&&\\ 									       									      									      
46 &   8& 1183 & &  8 10 17.868  & $-$49 26 14.62 &  17.850 &  1.847 &  0.527 &16.007  &15.502&  15.002 &M5.0&  $-$0.9 & $-$0.4 & & \llap{1}0 & $0.40\pm 0.04$  &2\\
47 &   8& 1274 & &  8 09 43.775  & $-$49 16 50.06 &  17.859 &  1.894 &  0.594 &15.964  &15.176&  15.021 &M6.0&  $-$2.2 & $-$3.1 & & \llap{1}3 & $0.38\pm 0.04$  &2\\
48 &  13& 1165 & &  8 10 41.234  & $-$49 12 22.11 &  17.864 &  1.836 &  0.569 &15.904  &15.211&  14.888 &M5.5&  $-$2.3 & $-$2.2 & & \llap{1}0 & $0.24\pm 0.06$  &2\\
49 &   9& 1595 & &  8 10 53.286  & $-$49 16 48.65 &  18.083 &  1.933 &  0.579 &16.010  &15.698&  15.120 &M5.0&  $-$1.7 & $-$4.0 & &  9 & $0.51\pm 0.05$  &2\\
50 &  18& 1480 &N&  8 09 53.162  & $-$49 05 37.98 &  18.088 &  1.731 &  0.515 &16.408  &15.511&  15.424 &M5.0& \llap{+1}3.4 & +4.4 &V& \llap{1}1 & \llap{$<$\,}0.08 &4\\
& & & & & & & & && & & & & &&&&\\ 									       									      									      
51 &  17& 1549 &N&  8 09 42.579  & $-$49 04 39.69 &  18.149 &  1.749 &  0.505 &16.419  &15.916&  15.722 &M5.0& \llap{+1}0.0 & \llap{+1}3.0 & &  7 & \llap{$<$\,}0.12 &4\\
52 &  13& 1850 & &  8 09 50.801  & $-$49 14 09.95 &  18.190 &  1.792 &  0.563 &        &      &         &M5.0&  $-$2.4 & +0.4 & &  8 & $0.52\pm 0.05$  &2\\
53 &   7& 1474 & &  8 09 08.079  & $-$49 21 59.71 &  18.243 &  1.929 &  0.583 &16.283  &15.278&  15.468 &M6.0&  $-$0.9 & $-$2.1 & &  9 & $0.47\pm 0.05$  &2\\
54 &  13& 1790 &N&  8 09 46.870  & $-$49 11 36.54 &  18.250 &  1.782 &  0.668 &16.390  &15.617&  15.377 &M5.0&  +1.7 & $-$0.9 & &  9 & $0.52\pm 0.05$  &2\\
55 &  13& 1776 & &  8 10 08.240  & $-$49 11 00.57 &  18.268 &  1.902 &  0.576 &16.244  &15.832&  15.705 &M5.0&  $-$0.9 & $-$0.6 & & \llap{1}0 & $0.40\pm 0.04$ &2\\
& & & & & & & & && & & &&&&&\\ 									       									      									      
56 &  13& 1828 & &  8 10 01.843  & $-$49 12 58.95 &  18.298 &  1.937 &  0.613 &16.323  &15.573&  15.588 &M5.5&  $-$1.5 &      & &  9 & $0.38\pm 0.06$\rlap{\,b} &2\\
57 &   7& 1425 & &  8 09 26.816  & $-$49 20 08.19 &  18.383 &  1.958 &  0.643 &16.365  &15.772&  15.498 &M5.5& $-$2.8 & $-$2.6 & &  8 & $0.71\pm 0.06$  &2\\
58 &   7& 1991 & &  8 09 47.146  & $-$49 26 33.24 &  18.564 &  2.088 &  0.674 &        &      &         &M6.5&  $-$1.5 & +3.2 & &  6 & $0.66\pm 0.07$  &2\\
59 &  12& 2131 & &  8 11 11.511  & $-$49 11 08.98 &  18.565 &  2.043 &  0.569 &16.520  &15.832&  14.586 &M6.0&  +2.6 & $-$5.2 & &  5 & $0.69\pm 0.09$  &2\\
60 &  14& 2399 &N&  8 09 04.030  & $-$49 18 43.54 &  18.633 &  1.976 &  0.716 &16.281  &15.889&  15.526 &M6.0&  +8.0 &      & &  7 & $0.94\pm 0.09$\rlap{\,b} &3\\
& & & & & & & & && & & & & &&&&\\ 									       									      									      
61 &   7& 1914 & &  8 09 46.706  & $-$49 23 24.75 &  18.805 &  1.964 &  0.625 &16.600  &15.975&  15.592 &M6.0&  $-$2.2 & $-$1.4 & &  5 & $0.72\pm 0.08$  &2\\
62 &   7& 2365 & &  8 09 34.569  & $-$49 23 53.49 &  18.964 &  2.054 &  0.551 &        &      &         &M6.0&  +0.5 & +7.5 & &  3 & $0.60\pm 0.14$  &2\\
63 &  12& 2833 & &  8 11 21.619  & $-$49 16 25.86 &  19.051 &  1.937 &  0.557 &        &      &         &M6.0&  \llap{+1}2.1 & \llap{+8}3.0 &V&  3 & \llap{$<$\,}0.28  &4\\
\hline								      
\end{tabular}
\end{table*}

\subsection{H$\alpha$ emission}

H$\alpha$ emission was seen in almost all our sky-subtracted spectra
and in principle could be used as an additional membership criterion
(e.g. Barrado y Navascu\'{e}s et al. 2004). However, whilst
chromospheric H$\alpha$ emission is expected from NGC 2547 members it
would also be expected from a large fraction of any possible field
M-dwarf contaminants with similar spectral types (Gizis, Reid \& Hawley
2002).  Unfortunately the presence and strength of H$\alpha$ emission
could not be accurately measured because the H$\alpha$ emission line
present in the sky spectrum is {\em much} stronger. The peak intensity
of the H$\alpha$ in the mean sky spectrum is 20-200 times the continuum
intensity around H$\alpha$ in our targets. Furthermore, the strength of
the sky H$\alpha$ emission varies by 20-30 per cent from fibre to fibre
because of the background H\,{\sc ii} region (see section 2.3),
resulting in effective EW errors of a few Angstroms even
in the brightest stars of our sample. Slit spectroscopy will be required to
obtain accurate estimates of chromospheric H$\alpha$ emission in these stars.

What we were able to do was search for extended H$\alpha$ emission
beyond the wings of the sky line. In particular, White \& Basri (2003)
have shown that emission lines with a full width at 10 per cent of
maximum of more than 270\,km\,s$^{-1}$ are indicative of ongoing 
accretion activity. Such objects would also typically have emission EWs
of more than 20\AA\ and hence extended emission wings easily visible in
our spectra. Following the method detailed in Kenyon et al. (2004) we
simulated a Gaussian H$\alpha$ emission line with an EW of 20\AA\ and
full width at 10 per cent of maximum of 270\,km\,s$^{-1}$ for each of
our spectra. No emission beyond this simulated profile (appropriately
broadened for those objects thought to be rapid rotators -- see section
3.3) was found. This result is not surprising. Jeffries et
al. (2000) found no accretion signatures in higher mass objects in NGC
2547 and a recent {\it Spitzer} survey of the cluster by Young et al. (2004)
finds less than 7 per cent of the low-mass cluster members have inner
dust discs betrayed by an infrared excess. Nevertheless, combined with
the precise LDB age derived for the cluster in this paper, NGC 2547
provides a useful upper limit to the lifetime of this early phase of
stellar evolution.

\subsection{Cluster membership}
\label{membership}

The measurements discussed above allow an assessment of cluster
membership. First, we assume that the 23 objects showing an Li feature
with an EW\,$>0.2$\AA\ and an RV consistent with cluster membership are
cluster members. The rationale is that curves of growth presented by
Zapatero-Osorio et al. (2002) suggest that the Li abundance is depleted
by about a factor of 100 or less in these objects (see
section~\ref{liabundance}), which is also the criterion we will adopt
for judging the location of the LDB. Field M dwarfs are extremely
unlikely to show an Li feature of this strength.

In addition there are two stars
which show Li at this level but have an RV inconsistent with cluster
membership. One of these (star 60) is probably a short period binary member of the
cluster, the other (star 5) is unlikely to be a cluster member based on
its colour and magnitude (see section~\ref{discussmembers}). 
The 9 objects which show no Li and
which have an RV inconsistent with cluster membership could be cluster
binaries though this is unlikely. Five of these objects are
fainter than our estimate of the LDB (see section~\ref{ldb}) and should
have exhibited Li if they were members and some of the others look
unlikely to be cluster members on the basis of their colours and
magnitudes (see section~\ref{discussmembers}). Until there
is further evidence we classify all these objects as
non-members.

Finally, there are 29 objects with an RV consistent with cluster
membership but showing no Li. These are very likely to be cluster
members, although given the $\sim 10$\kms\ range over which cluster
members are selected, it is possible that one or two contaminating
non-members remain in this subset. There are unlikely to be more than
this because we see no Li-poor, RV members {\it fainter} than the
location of the LDB. 

\section{Lithium abundances}
\label{liabundance}

\begin{figure}
\includegraphics[width=84mm]{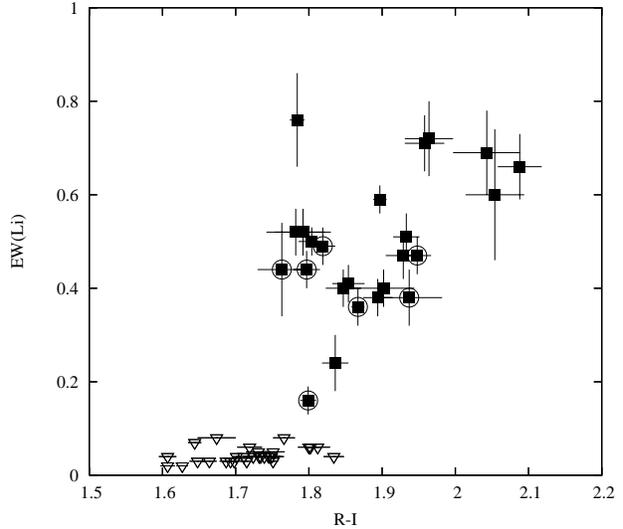}
\caption{Li EWs versus $R-I$ colour for Li-rich and Li-poor RV members
  of NGC 2547. Downward pointing triangles indicate 2-sigma upper limits.
  Encircled points are those with significant rotational broadening -- see section 3.3.}
\label{liplot1}
\end{figure}

\begin{figure}
\includegraphics[width=84mm]{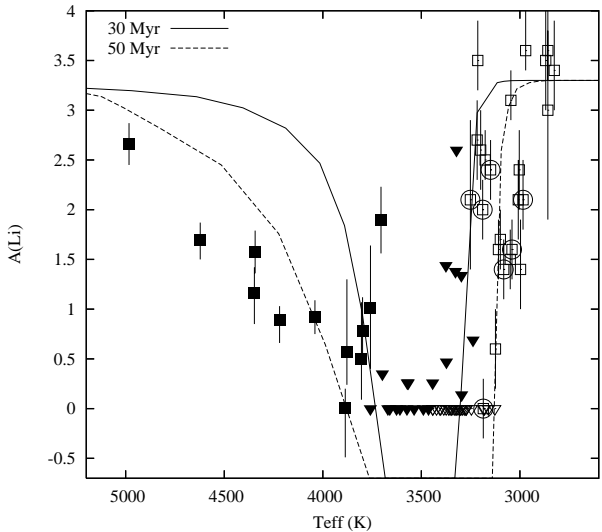}
\caption{Li abundances versus $T_{\rm eff}$ for RV members of NGC 2547
  from this paper (open symbols, encircled points indicate those with
  significant rotational broadening -- see section 3.3) 
  and from Jeffries et al. (2003 -
  filled symbols). Downward pointing triangles indicate 2-sigma upper
  limits. For comparison we show model predictions for the progression
  of Li depletion at 30 and 50\,Myr, taken from the models of Chabrier
  \& Baraffe (1997).}
\label{liplot2}
\end{figure}

We have estimated Li abundances, $A$(Li) (where $A$(Li) = $\log
[N$(Li)$/N$(H)]$+12$), for those objects which are RV members of the
cluster. We use the curves of growth for cool stars presented by
Zapatero-Osorio et al. (2002), and extended to $A$(Li)$=0.0$ by
Pavelenko \& Zapatero-Osorio (private communication), in order to
convert our EW values into Li abundances.  The difficulties in doing
this are described extensively in Jeffries et al. (2003), the main
problem being that the definition of the pseudo-continuum used to derive
the EW is not necessarily consistent between objects measured at
different spectral resolutions or with different degrees of rotational
broadening. It is also unknown to what extent there are model-dependent
systematic errors in the cool atmospheres used to determine the curves
of growth. For these reasons the {\em absolute} values of the Li
abundances should be treated with caution.

To estimate $T_{\rm eff}$ values for our objects we have used an
empirical relationship between intrinsic $R-I$ and $T_{\rm eff}$
defined by making a 120\,Myr isochrone from the Chabrier \& Baraffe
(1997) evolutionary models fit published Pleiades $RI$ data at a
distance of 132\,pc (see section 6.3.2 for details). This relationship
is shown in Fig.~\ref{riteff} and the $T_{\rm eff}$ values for the
cluster members (derived assuming that $E(R-I)=0.043$) are given in
Table~\ref{liabundancetable}. Statistical uncertainties which arise
from uncertainties in the $R-I$ photometry are small. However to allow
for some variability we assign uncertainties of $\pm 100$\,K for the
purpose of estimating possible Li abundance errors.  The Li abundances
were then obtained using a bicubic spline interpolation of the curves of
growth. Many of our EW upper limits imply Li abundances that fall {\em
well} below the lowest value in our grid, but we conservatively assign
an upper limit of $A$(Li)$<0.0$ to these. The statistical errors in the
abundances were found by perturbing the EW and $T_{\rm eff}$ by their
error bars and combining the two sources of error in quadrature. The EW
uncertainty is dominant because the curves of growth become nearly
independent of $T_{\rm eff}$ for $T_{\rm eff}<3500$\,K.  For this reason,
systematic uncertainties in the $T_{\rm eff}$ scale, which are quite
likely to be of order 100-200\,K at these cool temperatures, are not
important.  Li abundances and uncertainties are listed in
Table~\ref{liabundancetable}.

In Figs~\ref{liplot1} and~\ref{liplot2} we show plots of the Li EWs
versus $R-I$ colour and Li abundances versus $T_{\rm eff}$ for the NGC
2547 RV members. In the latter plot we have also included Li abundances
for higher mass objects in the cluster derived using the same curves of
growth (from Jeffries et al. 2003). A number of points arise from these
two figures.

(1) There is a clear pattern of undetectable Li for warmer stars
    ($R-I<1.75$). At slightly cooler
    temperatures there exist stars both with and without detectable Li
    and then for $R-I>1.84$ only Li-rich objects are found.

(2) The overall pattern of Li depletion agrees qualitatively (but not
    quantitatively) with those
    predicted by PMS evolutionary models (see also Fig.~9 and the
    discussion in Jeffries
    et al. 2003). In particular the sharp upturn at cooler temperatures
    - the LDB - is present. However it is impossible to define a
    precise or accurate age for NGC 2547 on the basis of
    Fig.~\ref{liplot2} because: (i) objects with and without Li
    co-exist at the crucial temperatures because of scatter caused by
    photometric errors, variability and unresolved binarity; (ii) the
    $T_{\rm eff}$ at the LDB {\em is} model dependent by of order $\pm
    100$\,K (see Table~\ref{ldbresults}), which translates to significant age uncertainties; (iii)
    there are systematic uncertainties of a similar size in the
    conversion from $R-I$ to $T_{\rm eff}$. These problems are solved
    in the next section by considering the data in the
    Hertzsprung-Russell diagram and finding the {\em luminosity} of the
    LDB.

(3) There appears to be a scatter in the Li abundances at low
    temperatures and even among the subset of objects with luminosities
    well below the LDB. This scatter arises directly from a spread in
    the EWs at a given colour. We believe that the spread in the EWs is
    real. Although some readers may question whether the approximation
    of equation~1 is strictly valid, it is unlikely to underestimate the
    EW errors by the factor of three that would be required to make
    them consistent with no spread. Systematic EW uncertainties are
    likely to be a monotonic function of $T_{\rm eff}$ and should not
    introduce scatter at a given colour. A similar Li EW scatter has
    been observed by Zapatero-Osorio et al. (2002) and Kenyon et
    al. (2004) for very low mass stars with $T_{\rm eff}<3500$\,K in the
    young $\sigma$~Orionis cluster and by Barrado y Navascu\'{e}s et
    al. (2004) in IC\,2391, but the spectra here are of higher quality and
    more importantly, taken at a homogeneous and relatively high resolution. 

    Whether the spread in abundances is
    real is a different matter. There is no theoretical explanation of
    such a scatter. All evolutionary models predict a sharp change from
    no Li to essentially cosmic Li over a small temperature
    range. There should be few, if any, transitional
    objects\footnote{Target number 21 with a weak but detectable Li
    line and $A$(Li)$=0.0\pm0.3$ may be an example of a transitional
    object, but could also be an example of an unresolved binary system
    with one component above and one component below the LDB.}.  Instead
    we think this problem may be related to our understanding of the
    upper atmosphere of very cool stars. Because the Li feature is
    saturated in these cool dwarfs it is very susceptible to NLTE
    effects or perhaps overionization caused by an overlying
    chromosphere (e.g. Houdebine \& Doyle 1995). This would cause a
    weakening of the Li feature in the most active stars and it is
    notable that: (i) the peak levels of Li abundance are roughly that
    expected for the cosmic Li abundance $A$(Li)$\simeq 3.3$; (ii) the
    rapid rotators tend to have lower Li EWs and abundances, although
    it would be difficult to discount problems with blending and
    continuum definition as an explanation for this. Given slit
    spectroscopy (to overcome the H$\alpha$ sky subtraction problems) or very
    deep X-ray observations it should be possible to look for an
    inverse correlation between Li strength and chromospheric/coronal
    activity.

(4) Despite the scatter, the objects do still clearly bifurcate between
    those objects with Li (with an EW$>0.2$\AA) and those with much
    lower upper limits. This division corresponds roughly to where 99
    per cent of the original Li has been depleted and we will use this
    to define the location of the LDB in the next section.

\begin{table}
\caption{Effective temperatures derived from the intrinsic $R-I$ colour
  and Li abundances (where $A$(Li)$=\log [N$(Li)$/N$(H)$]+12$) obtained
  from the curves of growth of Zapatero-Osorio et al. (2002) for RV
  members of NGC 2547 (status 1 and 2 objects from Table~\ref{main}). 
  Columns 1 and 4 are the identifiers used in Table~\ref{main}.}
\begin{tabular}{@{}rcc@{\hspace*{7mm}}rcc}
\hline
ID& $T_{\rm eff}$&$A$(Li)&ID& $T_{\rm eff}$&$A$(Li)\\
  & (K)          &       &   & (K) & \\
\hline
&&& & &\\
 1 &3272&$<0.0$                  &   31 &3149&$2.4^{+0.3}_{-0.3}$\\
 2 &3328&$<0.0$		   &   32 &3184&$<0.0$\\		
 4 &3423&$<0.0$		   &   34 &3374&$<0.0$\\		
 6 &3361&$<0.0$		   &   35 &3246&$<0.0$\\		
 7 &3291&$<0.0$		   &   36 &3160&$<0.0$\\		
 8 &3354&$<0.0$		   &   37 &3299&$<0.0$\\		
 9 &3386&$<0.0$		   &   38 &3322&$<0.0$\\		
10 &3442&$<0.0$		   &   39 &3189&$2.0^{+0.3}_{-0.3}$\\
11 &3348&$<0.0$		   &   41 &3099&$1.7^{+0.3}_{-0.3}$\\
12 &3127&$<0.0$		   &   42 &3214&$3.5^{+0.4}_{-0.3}$\\
13 &3402&$<0.0$		   &   44 &3251&$2.1^{+0.8}_{-0.7}$\\
14 &3282&$<0.0$		   &   45 &3176&$2.5^{+0.3}_{-0.2}$\\
15 &3345&$<0.0$		   &   46 &3109&$1.6^{+0.3}_{-0.3}$\\
16 &3406&$<0.0$		   &   47 &3049&$1.5^{+0.3}_{-0.3}$\\
17 &3442&$<0.0$		   &   48 &3124&$0.6^{+0.4}_{-0.4}$\\
18 &3180&$<0.0$		   &   49 &3003&$2.4^{+0.4}_{-0.4}$\\
19 &3333&$<0.0$		   &   52 &3199&$2.6^{+0.4}_{-0.4}$\\
20 &3303&$<0.0$		   &   53 &3008&$2.1^{+0.4}_{-0.4}$\\
21 &3186&$0.0^{+0.3}_{-0.3}$     &   54 &3217&$2.7^{+0.4}_{-0.4}$\\
22 &3272&$<0.0$		   &   55 &3040&$1.6^{+0.3}_{-0.3}$\\
25 &3275&$<0.0$		   &   56 &2997&$1.4^{+0.5}_{-0.4}$\\
26 &3046&$3.1^{+0.3}_{-0.2}$     &   57 &2971&$3.6^{+0.4}_{-0.2}$\\
27 &3082&$1.4^{+0.3}_{-0.3}$     &   58 &2826&$3.4^{+0.5}_{-0.4}$\\
28 &3304&$<0.0$		   &   59 &2869&$3.5^{+0.5}_{-0.5}$\\
29 &3314&$<0.0$		   &   61 &2963&$3.6^{+0.4}_{-0.3}$\\
30 &2984&$2.1^{+0.4}_{-0.3}$     &   62 &2858&$3.0^{+0.8}_{-1.1}$\\
&&&&&\\
\hline
\label{liabundancetable}
\end{tabular}
\end{table}

\section{The lithium depletion boundary age of NGC 2547}

\label{ldb}

\subsection{The location of the LDB}

\begin{figure}
\includegraphics[width=84mm]{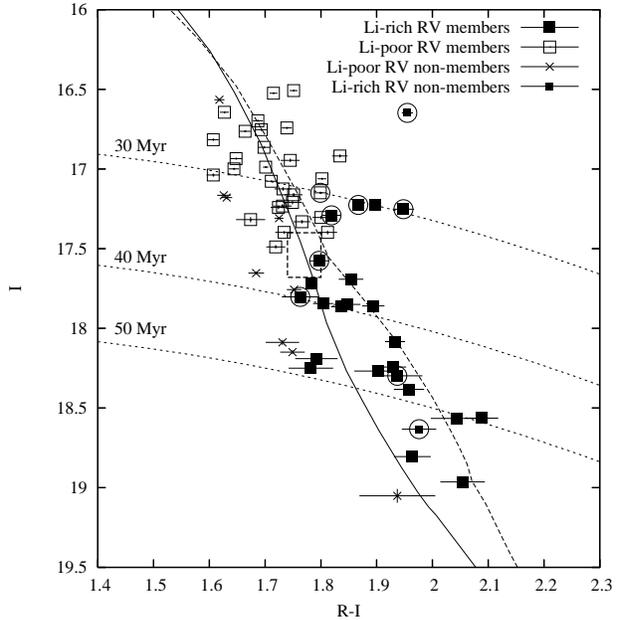}
\caption{I,R-I colour-magnitude diagram for our targets. 
Open squares represent objects with an RV consistent with cluster
membership but which have no Li, filled squares are those RV members
with Li, small filled squares are Li-rich stars with an RV discrepant
from the cluster and crosses represent objects with discrepant
velocities and no Li. Encircled points represent those objects with
discernible rotational broadening (see section 3.3).
The solid line is an empirically calibrated (see
text) 30\,Myr isochrone generated from the models of Baraffe et
al. (2002), the dashed line is a similar isochrone generated from the
D'Antona \& Mazzitelli (1997) models. We assume an intrinsic cluster
distance modulus of 8.10 and a reddening $E(R-I)=0.043$. The dashed box
indicates our estimate of the position and uncertainty of the LDB. The
dotted contours show isochrones (derived from the Chabrier \&
Baraffe 1997 models) corresponding to the location of the LDB in this diagram.}
\label{rildb}
\end{figure}

\begin{figure}
\includegraphics[width=84mm]{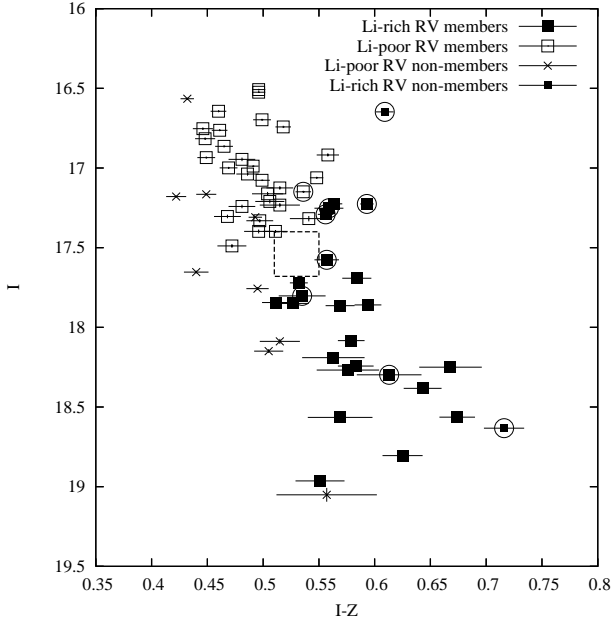}
\caption{I,I-Z colour-magnitude diagram. Symbols are as in
  Fig.~\ref{rildb}.}
\label{izldb}
\end{figure}

\begin{figure}
\includegraphics[width=84mm]{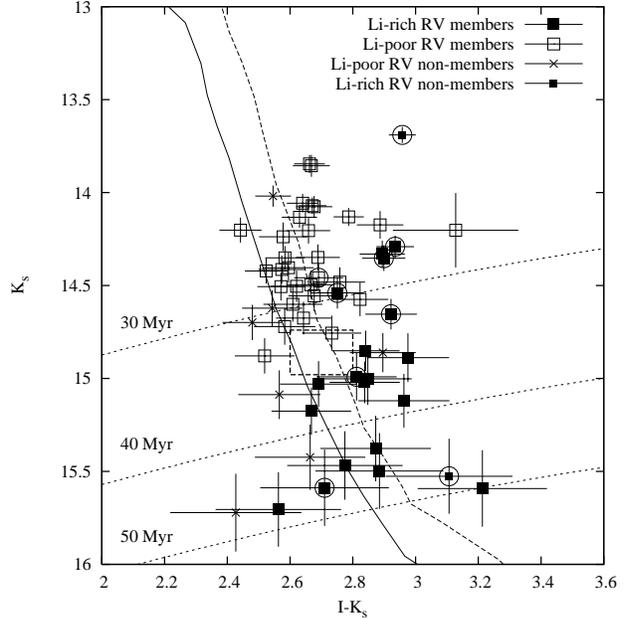}
\caption{$K_{s}$,$I-K_{s}$ colour-magnitude diagram. Symbols are as in
  Fig.~\ref{rildb}. The solid and dashed lines are 31.6\,Myr and
  20.0\,Myr isochrones with colours and magnitudes taken directly from
  the models of Baraffe et al. (2002) and assuming an intrinsic
  distance modulus of 8.10 and $E(I-K_{s})=0.092$. The dotted contours
  show isochrones (derived from the Chabrier \& Baraffe 1997 models)
  corresponding to the location of the LDB in this diagram.}
\label{ikldb}
\end{figure}

Figures~\ref{rildb}, \ref{izldb} and~\ref{ikldb} show the targets
plotted in the $I$ versus $R-I$, $I$ versus $I-Z$ and $K_{s}$ versus $I-K_{s}$
colour-magnitude diagrams (CMDs).  Objects classified as Li-rich
members, RV members with no Li, non-members on the basis of their RV or
Li-rich objects with discrepant RVs are represented with different
symbols. There is considerable scatter in these diagrams with no clear
PMS locus visible -- the isochronal ages discussed later in this paper
are derived from stars of higher mass (see section 6.3.2). 
There are probably several sources of this scatter
including binarity, rapid rotation, variability due to magnetic
activity and photometric uncertainties.

There is however a clear pattern among the Li abundances. The brighter
objects are Li-poor -- all RV cluster members with $I<17.22$ have no
Li. The fainter objects are Li-rich -- all RV cluster members with
$I>17.57$ have Li. However, there are both Li-poor and Li-rich objects
between these two limits which potentially confuse the location of the
LDB.  Three of the four Li-rich objects in this region have broad Li
lines suggesting rapid rotation which could result in them appearing
brighter and redder than suggested by a PMS isochrone for slowly
rotating stars (Pinsonneault et al. 1998). Alternatively their location
in all three CMDs suggests that these are unresolved binary systems,
which constitute roughly 30 per cent of the cluster population at these
colours (Jeffries et al. 2004). Perhaps a clearer way of looking at the
data is to note that while there may be several reasons why a star with
Li lies above the single star locus in the CMD, there are none (beyond
photometric error) that would cause a star without Li to lie below the
single star locus. The faintest object without Li has $I=17.49$ and two
others have $I=17.40$. This suggests that the LDB must be at $I\geq
17.4$. We contend that the LDB of NGC 2547 is at $I=17.54\pm0.14$ and
at colours of $R-I=1.77\pm0.03$ and $I-Z=0.53\pm0.02$. Similar
considerations lead us to place the LDB at $K_{s}=14.86\pm0.12$ and
$I-K_{s}=2.70\pm0.10$. These locations are represented by the boxes in
Figs.~\ref{rildb}, \ref{izldb} and~\ref{ikldb}.

\subsection{Age estimates}

The LDB location defined above can now be used to estimate the age of
NGC 2547. The approach we take is the same as that described in
Oliveira et al. (2003). Briefly, we use empirical relations between
bolometric correction (for $I$ or $K_{s}$) and colour ($R-I$ or
$I-K_{s}$), along with the appropriate distance modulus, extinction and
reddening (see section 1 and below) to
calculate the absolute $I$ or $K_{s}$
magnitude of the LDB at any age, where we assume that the LDB represents
the point at which 99 per cent of the initial Li has been depleted and
hence the EW of the \lii\ feature has dropped below about 0.2\AA. The age
estimate can be performed for different evolutionary models.
We note that our $K_{s}$ magnitudes are on the 2MASS system and were
converted to the CIT system (see Carpenter 2001) for the
purposes of these calculations.

Uncertainties in the derived age are produced by several factors (see
Jeffries \& Naylor 2001). The ones we include in our analysis are: (1)
The uncertainty in the position of the LDB in the CMD. These were
estimated in the previous section and probably represent uncertainties
of more than 1 sigma. (2) Systematic errors in the photometric
calibration. These are important for $I$, $R-I$ and
$I-K_{s}$. Jeffries et al. (2004) estimated external uncertainties for
stars with $R-I>1.5$ of up to $\pm 0.10$ mag in $I$ and $R-I$ and hence
there will be a systematic uncertainty of $\pm 0.10$ in the $I-K_{s}$
colours also. We assume a negligible external error for the $K_{s}$
magnitudes. (3) We assume an uncertainty in the reddening
$E(B-V)=0.06\pm0.02$ (Clari\'a 1982) and hence that $E(R-I)=0.043\pm0.014$,
$E(I-K_{s})=0.092\pm0.031$ and extinctions of $A_{I}=0.112\pm0.037$ and
$A_{K}=0.021\pm0.007$ (Rieke \& Lebofsky 1985).  (4) We assume an
uncertainty in the intrinsic distance modulus of $8.10\pm 0.10$ mag (see Jeffries \&
Tolley 1998; Naylor et al. 2002).

Table~\ref{ldbresults} list the results of these estimates for the
solar metallicity evolutionary models of Chabrier \& Baraffe (1997),
D'Antona \& Mazzitelli (1997), Siess, Dufour \& Forestini (2000) and Burke
et al. (2004). The uncertainties quoted include all the effects discussed
above, which are added in quadrature\footnote{This assumes the sources
  of uncertainty are independent. This is unlikely to be true for
  the distance modulus and reddening which are correlated in the sense
  that a larger assumed reddening leads to a larger distance modulus via
  main sequence fitting and both lead to a smaller LDB age.}.
Table~\ref{ldbresults} also gives the bolometric magnitude of
the LDB, and the masses and $T_{\rm eff}$ at the LDB for each model.

As a comparison, we have also recalculated the LDB age of IC 2391 using
our models and bolometric corrections.  The LDB location was found by
Barrado y Navascu\'{e}s et al. (2004) to be at $I=16.22\pm0.10$,
$R-I=1.90\pm0.05$ and $K_{s}=13.49\pm0.10$, $I-K_{s}=2.75\pm 0.10$. An
intrinsic distance modulus of $5.95\pm 0.10$ and reddening equal 
to that of NGC 2547 were also assumed. The LDB ages and parameters
for IC 2391 are also given in Table~\ref{ldbresults}.

In addition to the random error bars quoted on the ages in
Table~\ref{ldbresults}, there are other small systematic uncertainties
affecting the LDB age scale as a whole -- the gravity dependence of the
bolometric correction and the exact level of Li depletion that the LDB
corresponds to (see Burke et al. 2004).  The theoretical models of
Baraffe et al. (2002) yield $I$-band bolometric corrections as a
function of colour and age (and hence gravity). It seems that the
relationship between bolometric correction and $R-I$ is more age
sensitive, with bolometric corrections lower by about 0.05 mag for
stars at 30\,Myr compared with 5\,Gyr. The effect is limited to about
0.02 mag in the $I-K_{s}$ relation. Because the bolometric corrections we
use are derived from high gravity field dwarfs a small adjustment to the
bolometric corrections may be required. However, changes of this size
would lead to LDB ages that were younger by only $\sim 1$\,Myr.
Similarly, Li depletion is so rapid that whether the LDB corresponds to
90 per cent depletion or 99.9 per cent depletion changes the LDB ages
by only $\pm 1$\,Myr.

The ages in Table~\ref{ldbresults} show very close agreement both in
terms of the ages using bolometric corrections deduced from the $R-I$
and $I-K_{s}$ colours and between ages derived from different models
incorporating different physics. Taking the average result from both
colours, the age of NGC 2547 ranges from 34\,Myr to 36\,Myr and the
age of IC 2391 from 48\,Myr to 53\,Myr. The youngest ages in both cases
are given by the D'Antona \& Mazzitelli (1997) models. The difference in LDB
age between IC 2391 and NGC 2547 is very consistent at 14-17\,Myr.

As a final estimate of the LDB age we compared the absolute $I$ and $K_{s}$
magnitudes at the LDB directly with the models of Baraffe et
al. (2002), remembering to convert the 2MASS magnitudes to the CIT
system predicted by the models.  In other words we are using {\it
theoretical} bolometric corrections produced by the cool model atmospheres.
The results for NGC 2547 and IC 2391 
are listed in the last column of
Table~\ref{ldbresults} and are very close to the estimates based on
empirical bolometric corrections and the same evolutionary models
(those of Chabrier \& Baraffe 1997).

\begin{table*}
\caption{ The LDB ages for NGC 2547 and IC 2391 using the $I$ vs $R-I$
  and $K$ vs $I-K$ CMDs using different models and a summary of model parameters
  at the LDB. }
\begin{tabular}{llccccc}
\hline
 &  & Chabrier \& & D'Antona \& & Siess et al. & Burke
 et al. & Baraffe et al.\\
 &     & Baraffe (1997)    & Mazzitelli (1997)  & (2000) $z=0.02$  &
 (2004) & (2002) theoretical \\
\hline
{\bf NGC 2547} & & & &&&\\
&&&&&&\\
LDB ages & $I$ vs $R-I$ & $35.4\pm3.3$ & $34.0\pm3.7$ & $36.2\pm3.4$ & 
 $36.1\pm4.4$ & $36.3\pm2.8$ \\
(Myr)         & $K_{s}$ vs $I-K_{s}$ & $34.4\pm2.7$ & $33.8\pm2.8$ & $35.4\pm2.6$ &
 $35.4\pm3.1$ &$35.0\pm2.6$ \\
&&&&&&\\
Parameters &$M_{bol}$$^{1}$ & $9.58\pm0.16$ &$9.58\pm0.16$ &$9.58\pm0.16$&$9.58\pm0.16$&$9.63\pm0.16$\\
at the LDB &Mass$^{1}$ ($M_{\odot}$) & $0.17\pm0.02$ &$0.17\pm0.02$&$0.15\pm0.02$ &$0.17\pm0.01$&$0.17\pm0.02$\\
           &$T_{\rm eff}$$^{1}$ (K)&$3250\pm 30$ &$3140\pm 20$ &$3230\pm 40$&$3240\pm 40$&$3240\pm 30$\\
&&&&&&\\
{\bf IC 2391}&&&&&&\\
&&&&&&\\
LDB ages & $I$ vs $R-I$ &
 $49.1\pm4.9$&$47.3\pm4.3$&$49.6\pm4.8$&$52.5\pm5.7$ &
 $50.8\pm 3.9$ \\
(Myr) &$K_{s}$ vs $I-K_{s}$
 &$50.4\pm3.8$&$48.5\pm3.2$&$50.5\pm3.5$&$54.1\pm4.4$ &$49.6\pm3.8$\\
&&&&&&\\
Parameters &$M_{bol}$$^{1}$&$10.34\pm 0.14$ &$10.34\pm 0.14$ &$10.34\pm 0.14$&$10.34\pm 0.14$&$10.36\pm0.14$\\
at the LDB &Mass$^{1}$ ($M_{\odot}$)&  $0.12\pm 0.01$&$0.12\pm 0.01$ &$0.11\pm0.01$ &$0.12\pm0.01$&$0.12\pm0.01$\\
&$T_{\rm eff}$$^{1}$ (K)&$3140\pm 30$ &$3070\pm 20$ &$3100\pm 30$&$3150\pm 40$&$3130\pm 30$\\
\hline
\multicolumn{7}{l}{1 - these are averages from the very similar results for the two diagrams}\\
\end{tabular}
\label{ldbresults}
\end{table*}

\section{Discussion}

\subsection{Comparison with Oliveira et al. (2003)}

The analysis of Oliveira et al. (2003) arrived at a slightly older
lower limit to the LDB age for NGC 2547 than we have found in this
paper. Oliveira et al. found no evidence for a population of Li-rich
objects with $I<17.2$, although they did find a few examples of Li
detections of brighter stars which might be either binary members of
the cluster or perhaps not members of the cluster at all -- see below.
At fainter magnitudes there were a few tentative Li detections, but
to place limits on the location of the LDB, Oliveira et al. averaged the
spectra of targets with $17.3<I<17.8$ and with $I\geq 17.8$ separately.
Evidence for an Li line was found in the average spectrum of the
fainter sample, but not the brighter. Hence Oliveira et al. placed the
LDB at $17.8<I<18.3$ and derived an age at least 3\,Myr older than in
this paper.

With the benefit of hindsight and in the light of the {\it much} better
data obtained with VLT/GIRAFFE we can explain the probable cause of
this discrepancy. We have found that the LDB is at $I=17.54\pm 0.14$, but
5 of the 10 objects in the $17.3<I<17.8$ sample of Oliveira et al. are
brighter than this and hence unlikely to be Li rich. In addition we see
from Fig.~\ref{rildb} that some level of contamination of the sample
is also likely, such that perhaps 1-2 of the 5 objects with
$17.54<I<17.8$ might be non-members and hence
Li-poor. Given this, it is not hard to see why Li may not have been
detected in the average spectrum of this subsample. On the other hand,
the fainter sample should all be Li-rich , apart from
contamination by Li-poor non-members at the level of $\leq 25$ per
cent, and hence an Li detection would be expected.

There are also some discrepancies between the strengths of Li detected
in this paper and those measured in Oliveira et al. (2003).  We have
observed three of the stars claimed to be Li-rich by Oliveira et al. Of
these, we also detect Li in star 31 ($=$ Oliveira no.75, O75) and star 42 ($=$
O91). But, we find a very low upper limit to the Li EW of star
50 ($=$ O100), where Oliveira et al. claim a significant
detection.  We have also observed six stars with a significant Li
detection -- stars 5($=$O46), 27($=$O73), 39($=$O85), 41($=$O89), 45($=$92)
and 53($=$O106) for which Oliveira et al. found upper limits to the Li
EW. Of these, two have a detection consistent with the previous upper
limit (stars 27 and 53), but the other four have Li EWs greater than
the previous 2-sigma upper limit. Three of these (stars 39, 41 and 45)
were classed as non-members by Oliveira et al. on the basis that their
narrow band spectroscopic indices (the same as those used in this paper)
were too low for stars of their colour. It now seems clear that the
``non-members'' found by Oliveira et al. were almost certainly
instances of under-subtracted sky, because even the contaminants in the
photometrically selected cluster candidates have spectroscopic indices
similar to those of cluster members. An under-subtracted sky will
result in lower spectroscopic indices, dilution of any Li EW and an
erroneously optimistic assessment of the SNR of the spectrum.  Hence it
is not surprising that stars classed as non-members in Oliveira et
al. do not have Li detections.

\subsection{Cluster membership and contamination}
\label{discussmembers}

\begin{figure}
\includegraphics[width=84mm]{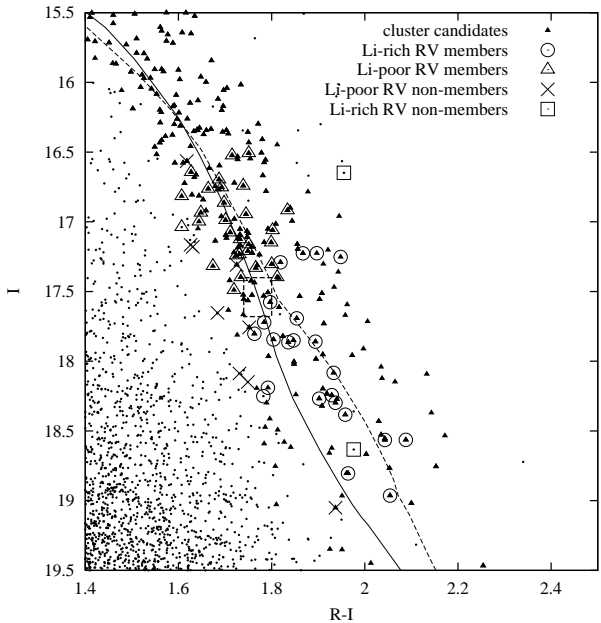}
\caption{I,R-I colour-magnitude diagram for our targets superimposed on
  the general population towards NGC 2547, showing those objects
  identified as photometric cluster candidates by Jeffries et
  al. (2004). The solid line shows the empirically calibrated 30\,Myr
  isochrones used to select members, which were generated from the
  models of Chabrier \& Baraffe (1997 -- solid line) and D'Antona \&
  Mazzitelli (1997 -- dashed line). Photometric selection also made use
  of the $I$ vs $I-Z$ and $R-I$ vs $I-Z$ diagrams.
}
\label{memberplot}
\end{figure}

The work of Jeffries et al. (2004) in determining the mass function of
NGC 2547 relied solely on photometric criteria to select cluster
candidates. The spectroscopic results presented here offer an
opportunity to check those photometric criteria and to estimate the
amount of contamination present in a photometrically selected sample.

Figure~\ref{memberplot} is an $I$ versus $R-I$ CMD showing 
the catalogue of stars from which
Jeffries et al. (2004) selected cluster candidates, their photometrically
selected cluster candidates and symbols representing the status of
stars that are analysed in this paper. Note that the cluster candidates
were also selected on the basis of their positions in the $I$ versus
$I-Z$ and $R-I$ versus $I-Z$ diagrams (see Jeffries at al. for
details). There are several points arising from Fig.~\ref{memberplot}.

(1) Looking at the distribution of RV selected members and RV non-members from this paper
    and comparing them with the distribution of photometrically selected
    candidates it appears that the photometric selection criteria
    employed by Jeffries et al. (2004) were quite appropriate.
    The presence of a ``line'' of five RV non-members {\it just below} the lower
    boundary of the cluster candidate distribution suggests that this
    boundary is well chosen, although we have also
    found two RV members (one Li-rich) near this boundary which were
    not photometrically selected as cluster candidates. 

(2) The upper
    boundary of the photometric selection is less well explored by our
    spectroscopic data. However
    this has lesser implications for the completeness and contamination
    of the photometrically selected sample because the density of stars in the CMD
    decreases rapidly towards brighter magnitudes.

(3) Looking at spectroscopically observed stars which are also
    photometrically selected candidates we find four RV non-members
    (plus another RV non-member which is Li-rich and a probable cluster
    binary -- star 60). In contrast we have found 50 RV members among
    the photometrically selected cluster candidates of which perhaps
    only 1-2 are random contaminants (none of which will be Li-rich).
    Hence the level of contamination among the photometrically selected
    cluster candidates is only about 10-15 per cent for the central
    regions of the cluster in the magnitude interval
    $16.5<I<19$. Given that we have observed nearly all the cluster candidates
    within the VLT/GIRAFFE field of view, this translates into a
    spatial density of about $30\pm 15$ contaminants per square degree in this
    magnitude range, or $45\pm 18$ if we assume there are an additional
    2 contaminants among the Li-poor RV members. This compares
    reasonably well, though is a little lower 
    than the estimates from Jeffries et al. (2004) of 80 per square
    degree based on a population synthesis or the $64\pm 26$ per square
    degree based on the number of photometric cluster candidates found
    in regions offset from the cluster. Differences in contamination
    levels of this size would
    have no significant effect on the mass and luminosity functions
    for NGC 2547 determined by Jeffries et al. (2004).

(4) There is scope to improve the precision of the LDB location
    in the CMDs. There are a further 8 photometric cluster candidates that
    we have not observed that lie {\it inside} the LDB error box in the
    $I$ versus $R-I$ CMD. However, the contributions to the error budget
    from uncertainties in the cluster distance and uncertain
    photometric calibrations for very cool stars play an equal role in
    the precision of the LDB age estimate.

(5) Star 5 is Li-rich but quite clearly not a photometric member of the
    cluster in any of the CMDs and not an RV member either. This star
    is too far above the cluster locus to be a a binary member. A
    single star of this colour would have to be younger than 50\,Myr to
    have preserved its Li and such stars should be quite rare among the
    general field population. One possibility is that star 5 is a
    low-mass member of the Vela OB2 association which is sparsely
    spread over a $\sim 10$ degree diameter and which encompasses NGC 2547.
    Vela~OB2 is at a distance of $\sim 400$\,pc (de Zeeuw et al. 1999) and has
    an age of a few Myr. A population of low-mass, Li-rich PMS
    stars has been found in this association by Pozzo et al. (2000,
    2001). If star 5 were at a distance of 400\,pc, then its position
    in the CMDs suggests an age of about 4\,Myr (Baraffe et al. 2002), in
    excellent agreement with this hypothesis.
    Furthermore, star 5 has an RV which is 7.3\kms\ larger than NGC
    2547 average, which itself has a mean heliocentric RV of $+12.8\pm 1.0$\kms\ (Jeffries
    et al. 2000). Thus the heliocentric RV of star 5 is $20.1\pm
    1.0$\kms, very close to the heliocentric RV of
    18\kms\ obtained for the Vela OB2 PMS population by Pozzo et al. (2001).
A similar explanation is likely for the Li-rich star RX\,2 found in the
    NGC 2547 study of Jeffries et al. (2003). It lies 2 mag above the
    NGC 2547 PMS locus, exhibits undepleted Li and has an RV consistent
    with either NGC 2547 or Vela OB2.    
It is fortunate for membership studies in NGC 2547 that the
    distance/age combination of Vela OB2 places its low-mass
    population well above the NGC 2547 locus in CMDs.

\subsection{Comparison of LDB ages with isochronal ages}

The LDB ages determined for NGC 2547 and IC 2391 in
Table~\ref{ldbresults} (and also the Alpha Per and Pleiades clusters --
see Jeffries \& Naylor 2001) are very precise thanks to the steep
luminosity dependence of the age at which Li is burned in
fully convective stars. These LDB ages are also likely to be very accurate
in an absolute sense. The range of models we have employed
use a wide variety of equations of state, radiative opacities,
atmospheres and convective treatments and yet the difference between
the largest and smallest age estimate in the case of NGC 2547 is only
1.9\,Myr! Of course there is always a concern that some physics
neglected by all the models may be important, such as rotation or
the presence of supporting magnetic fields in the convection
zone. Preliminary work suggests that the effects of these will be to
increase the LDB ages, but only by a very small amount (Burke et
al. 2004; F. D'Antona private communication).
As the LDB ages are both precise and accurate they offer the
opportunity to test the less well constrained physics that controls age
determinations from higher mass stars. 

\subsubsection{Nuclear turn off ages}
The LDB results for IC 2391, the
Alpha Per and Pleiades clusters have shown that nuclear turn-off ages
deduced from high mass stellar models without convective core
overshooting are younger by factors of 1.5 than LDB ages (Barrado y
Navascu\'{e}s et al. 2004). Unfortunately NGC 2547 may or may not be
consistent with this picture. The nuclear turn-off age of NGC 2547 is
ill-determined ($55\pm25$\,Myr based on models with a small amount of
convective overshoot -- Jeffries \& Tolley 1998), mainly because it
hinges on the photometry and reddening of just one star (HD 68478, not HD
68468 as stated in Jeffries \& Tolley 1998).

\subsubsection{Empirical low-mass isochrones in colour-magnitude diagrams}

\begin{figure}
\includegraphics[width=84mm]{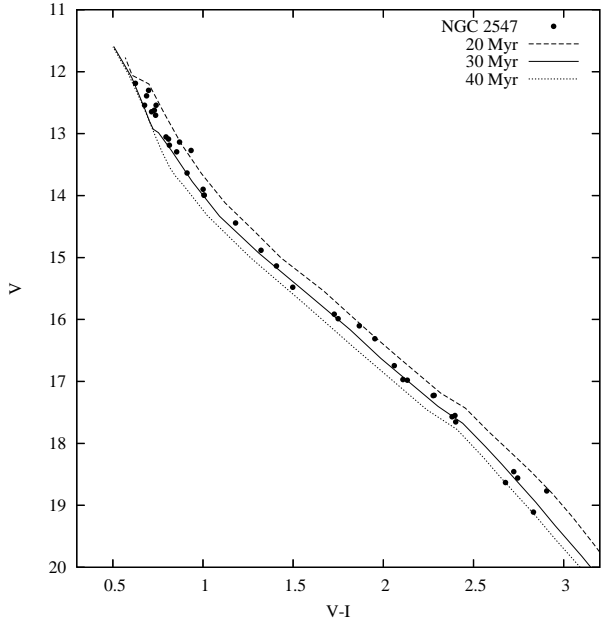}
\caption{An example $V$ versus $V-I$ CMD for presumed single, spectroscopically
  confirmed members of NGC 2547 (spots). Empirically calibrated 
  isochrones at 20, 30 and 40\,Myr
  are shown, derived from the theoretical models of Baraffe
  et al. (2002, with a mixing length parameter of 1.9 pressure scale
  heights) with a distance, reddening and extinction appropriate for
  NGC 2547.
}
\label{viempirical}
 \end{figure}

\begin{table}
\caption{Empirical isochronal ages (in Myr) for NGC 2547}
\begin{tabular}{lcc}
Evolutionary    &  \multicolumn{2}{c}{CMD} \\
Model           &  $V$ vs $V-I$ & $I$ vs $R-I$\\
\hline
&& \\
D'Antona \& Mazzitelli  & 24 & 26 \\
(1997)    &                  &    \\
Baraffe et al. (2002)   & 29 & 31 \\
(ML$=1.9$ scale heights)&    &    \\
Baraffe et al. (2002)   & 34 & 36 \\
(ML$=1.0$ scale heights &    &    \\
Siess et al. (2002)     & 26 & 28 \\
$z=0.02$
&& \\
\hline
\end{tabular}
\label{isochroneage}
\end{table}

The ages of young clusters can also be derived rather precisely (but
not necessarily accurately -- see below) by
fitting low-mass isochrones to cluster members in CMDs.
The rate of descent towards the ZAMS is mass dependent but
also depends somewhat on the treatment of convection and the
details of the stellar atmospheres. Here NGC 2547 is an excellent
cluster to check the consistencies of the age determinations. It has a
well studied lower main sequence with very precise photometry and a
well defined PMS locus.  Unfortunately the absolute accuracy of the
method is also dependent on how the model temperatures and
luminosities are converted into colours and magnitudes before
comparison with the data (or vice-versa).

In previous papers (e.g. Naylor et al. 2002; Oliveira et al. 2003;
Jeffries et al. 2004) a method was described for empirically
deriving the relationship between colour and $T_{\rm eff}$, using the
well-observed Pleiades cluster as a calibrator. We assume the Pleiades
distance and age are 132\,pc and 120\,Myr (roughly the LDB age) and
adopt the same relationships between bolometric correction and
colour as were used in deriving the LDB ages.  Then,
assuming the same colour-$T_{\rm eff}$ relation applies to clusters as young as
NGC 2547 (which is not contradicted by theoretical atmosphere models),
isochrones can be generated to find a cluster age. The method is
insensitive to the age assumed for the Pleiades and changing the
Pleiades distance simply changes the distance to the fitted cluster by a
similar factor (and hence the LDB age -- see the discussion in Oliveira et al. 2003).

Figure~\ref{rildb} shows examples using the
D'Antona \& Mazzitelli (1997) and Baraffe et al. (2002, with a mixing
length parameter of 1.9 pressure scale heights) models for an age of 30\,Myr. Other
examples of fits to NGC 2547 and IC 2391 covering stars at higher
masses and in other CMDs can be found in Naylor et al. (2002), Oliveira
et al. (2003) and Jeffries et al. (2004). We have uniformly reanalysed
the $V$ versus $V-I$ and $I$ versus $R-I$ CMDs for NGC 2547 to estimate
low-mass isochronal ages using the same models used to estimate
the LDB ages. We have taken
spectroscopically confirmed stars from Jeffries et al. (2000, 2003) and
excluded those stars which were considered to be binary systems in
an attempt to define a clean single star isochrone.
For stars with $1.0<V-I<2.5$ and $0.5<R-I<1.5$, spectroscopically
selected members of NGC 2547 lie on a very well defined PMS locus in the $V$
versus $V-I$ and $I$ versus $R-I$ CMDs (e.g. see
Figs.~\ref{viempirical} and~\ref{iritheory}). Warmer stars have already
reached the ZAMS and offer no age discrimination. Cooler stars appear
to show a more scattered PMS (see Figs.~\ref{rildb}
and~\ref{iritheory}), due to a combination of variability and
photometric errors, but may also suffer from systematic
photometric calibration uncertainties.

Model isochrones were generated using the Pleiades as an empirical
calibrator of the colour-$T_{\rm eff}$ relation and then matched to the
NGC 2547 data using an intrinsic distance modulus of 8.1,
(which is constrained by higher mass stars that have already reached the
ZAMS) reddenings of $E(V-I)=0.077$ and $E(R-I)=0.043$ and the
corresponding extinction values. The results are given in
Table~\ref{isochroneage} and an illustrative example of the isochrone
matching shown in Fig.~\ref{viempirical}. The ages deduced from the two
CMDs are very consistent, but vary from about 25\,Myr for the D'Antona
\& Mazzitelli (1997) models (a value independently arrived at by
Stauffer et al. [2003] using a similar technique), up to about 35\,Myr for Baraffe et
al. (2002) models with a mixing length set to 1.0 pressure scale heights.
The photometry available for the IC\,2391 cluster is more scattered,
but suggests an isochronal age that is 5 to 10\,Myr older using any of
these models.

The relative isochronal ages from different models are
very precise because the alternative isochrones are almost
parallel in the colour ranges considered. There are however uncertainties
which shift the ages from all models approximately equally.
The main sources of experimental error in this technique are the
uncertainties in the adopted distance modulus and reddening of the
cluster. Assuming the same values used to derive the LDB ages we find
$\mp 5$\,Myr for a $\pm 0.1$\,mag change in distance modulus. This is a
change in the same sense, but larger in magnitude than for the LDB age,
where we find about $\mp 1.6$\,Myr for a $\pm 0.1$ mag change in the
distance modulus of NGC 2547. A $\pm 0.02$ mag change in $E(B-V)$
results in a change in the derived intrinsic distance modulus (from the
ZAMS stars) of about $\sim \pm 0.1$, but does not significantly change
the isochronal age estimates from the lower mass stars since the
reddening vector is close to parallel with the PMS isochrones in the
CMDs. Statistical errors in deciding which isochrone fits best are
only about $\pm 2$\,Myr for NGC 2547 where, for $1.0<V-I<2.5$ and
$0.5<R-I<1.5$ the stars lie almost like ``beads on a string''. The scatter is greater 
at redder colours in NGC 2547 and also in IC 2391 at all colours, where an extra
$\pm 5$\,Myr uncertainty in the isochronal age is warranted.

The main source of systematic error is the compatibility of the
photometric calibrations of the Pleiades calibration data and the
younger cluster data and the possibility that the colour-$T_{\rm eff}$
relation changes for stars with lower gravity.  These factors are much
less important for stars at 30\,Myr with $R-I<1.5$ and $V-I<2.5$, which
have gravities quite similar to Pleiades stars of the same colour and
which are calibrated by many photometric standards of appropriate
colour. We attach less weight to empirical isochrone comparisons for
cooler stars, where gravities are lower than in the Pleiades, where the
photometric calibration may be uncertain by $\pm 0.1$ and where there
appears to be more scatter in the photometry of genuine cluster
members. However, even with these uncertainties it is encouraging that
the empirical 30\,Myr isochrones shown in Fig.~\ref{rildb} are
certainly compatible with the data for the cooler stars observed in
this paper.  

Stauffer et al. (2003) have shown that the deduced
isochronal age may also depend to a certain extent upon which colour indices
are used. In particular they find that young, magnetically active stars
suffer from a ``blue excess'' with respect to older stars. This would make
ages determined from the $V$ versus $B-V$ CMD problematic. However,
they also show that this problem is much less serious using the $V$
versus $V-I$ CMD upon which our empirical isochronal ages for NGC 2547
largely rest. In any case, by using the Pleiades as an empirical
calibrator of the colour-$T_{\rm eff}$ relationship we potentially nullify
these problems because the cool stars of NGC 2547 have very similar
magnetic activity levels to those in the Pleiades (Jeffries \& Tolley
1998). We also find that isochronal ages determined from the $V$ versus
$V-I$ and $I$ versus $R-I$ CMDs are entirely consistent. The same may
not be true when it comes to comparing theoretical isochrones with the
NGC 2547 data, because the best theoretical atmospheres currently
contain no contribution from chromospheres, spots or plages (see
section 6.3.4).

The LDB age for IC 2391 is larger than its low-mass isochronal age when
determined using the same models,
although the discrepancy is perhaps not very significant at about $12\pm
8$\,Myr.  It was partly this and partly the overestimate of the LDB age
for NGC 2547 that led Oliveira et al. (2003) to speculate that LDB ages
were {\it systematically higher} than isochronal ages in low-mass stars,
perhaps indicating the need for new physics in the low-mass
evolutionary models. Our new and precisely determined LDB age for NGC
2547 is only larger by a few Myr than the isochronal ages. The largest
discrepancy occurs for the D'Antona \& Mazzitelli (1997) models, and is
only $10\pm 6$\,Myr. Discrepancies of this size could reasonably be
accounted for by a small decrease in the cluster 
distance modulus of about its estimated error bar or could be due to
small systematic changes in the colour-$T_{\rm eff}$ relationship with age.

\subsubsection{The empirical temperature scale}
\begin{figure}
\includegraphics[width=84mm]{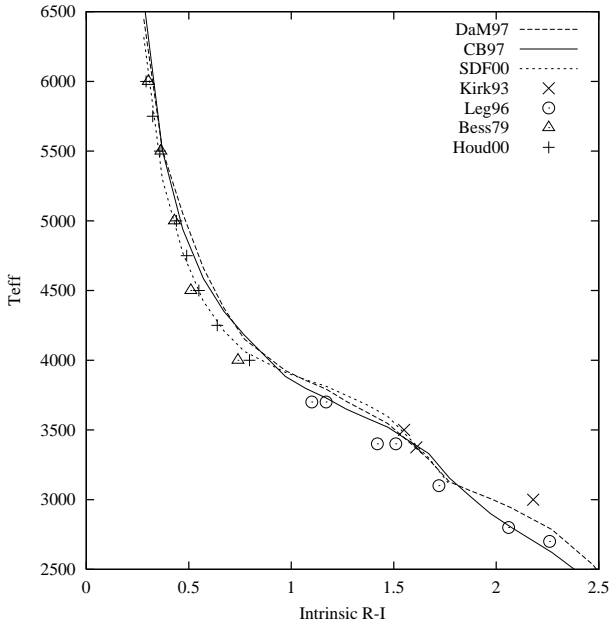}
\caption{The relationship between intrinsic $R-I$ and $T_{\rm eff}$
  derived by requiring evolutionary models to match Pleiades data at an
  age of 120\,Myr and distance of 136\,pc. The relationships for the
  Chabrier \& Baraffe (1997 - CB97, with overshooting of 1.9 pressure scale
  heights), D'Antona \& Mazzitelli (1997 - DaM97) and Siess et al. (2000
  - SDF00, solar metallicity model) are shown. Also shown are the
  semi-empirical relationships found by Houdashelt et al. (2000 -
  Houd00) and Bessell (1979 - Bess79) and the empirical determinations
  of Kirkpatrick et al. (1993 - Kirk93) and Leggett (1996 - Leg96).
}
\label{riteff}
 \end{figure}

The near consistency achieved between the LDB and isochronal ages is
almost independent of the choice of model.  This insensitivity to the
chosen model arises from the intrinsic similarity of the isochrones
generated by the various models at this age, but also by our
calibration technique of fixing the age of the Pleiades and hence
deriving a slightly different relationship between colour and $T_{\rm
eff}$ for each model. There {\it are} some differences in the isochrone
shapes for cool young stars (see for instance Fig.~\ref{rildb}), but
unfortunately these occur at colours where uncertainties in the present
photometric calibrations make it impossible to judge which models best
describe the data.

In principle we could compare the required intrinsic colour-$T_{\rm
eff}$ relationship with empirical and semi-empirical determinations of
$T_{\rm eff}$ to see which model correctly describes this relationship.
Figure~\ref{riteff} shows this comparison for $R-I$ versus $T_{\rm
  eff}$, 
the plot for $V-I$ versus $T_{\rm eff}$ is qualitatively very
similar.  Given the uncertainties in the empirical $T_{\rm eff}$
determinations -- typically 200\,K and the possible $\pm 0.1$ mag
uncertainties in the photometric colours of very red stars we would be
hard pressed to choose between the models in this way.

Finally, we can compare the $T_{\rm eff}$ at which the models predict
the LDB (listed in Table~\ref{ldbresults}) with the $T_{\rm eff}$ 
corresponding to the colour of the LDB from
Fig.~\ref{riteff}. The $T_{\rm eff}$ at the LDB ranges from 3250\,K,
for the Chabrier \& Baraffe (1997) models (which use a detailed model
atmosphere) to 3140\,K for the D'Antona \& Mazzitelli (1997) models
(which use a grey atmosphere). The derived {\em intrinsic}
($R-I$)-$T_{\rm eff}$ relations in Fig.~\ref{riteff} give very similar
$T_{\rm eff}$ values for all the models of 3185-3220\,K at
$(R-I)_{0}=1.73$. Given the possible 0.1 mag uncertainties in $R-I$,
corresponding to $T_{\rm eff}$ uncertainties of $\sim 150$\,K, all the
models are satisfactory in the sense that the isochronal $T_{\rm eff}$
at the colour of the LDB and the theoretical $T_{\rm eff}$ at the LDB
are consistent.

\subsubsection{Comparison with theoretical low-mass isochrones}

\begin{figure}
\includegraphics[width=84mm]{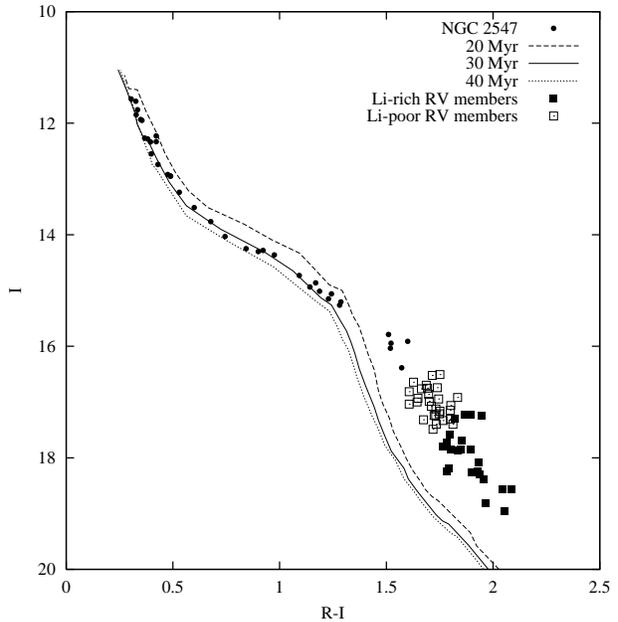}
\caption{An $I$ versus $R-I$ CMD for presumed single, spectroscopically
  confirmed members of NGC 2547 (spots) and the radial velocity members
  identified in this paper (squares). Isochrones at 20, 30 and 40\,Myr
  are shown and taken directly from the theoretical models of Baraffe
  et al. (2002, with a mixing length parameter of 1.9 pressure scale
  heights) with a distance, reddening and extinction appropriate for
  NGC 2547.
}
\label{iritheory}
 \end{figure}
\begin{figure}
\includegraphics[width=84mm]{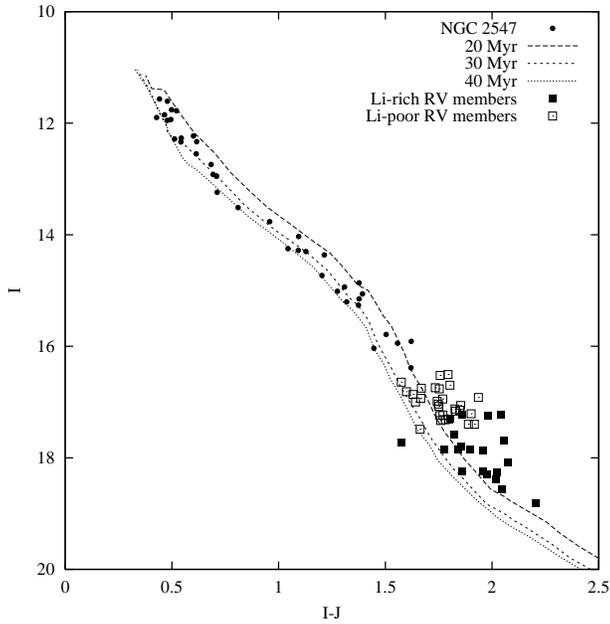}
\caption{An $I$ versus $I-J$ (2MASS system) CMD for NGC 2547. Symbols and lines are as
  for Fig.~\ref{iritheory}
}
\label{iijtheory}
 \end{figure}
\begin{figure}
\includegraphics[width=84mm]{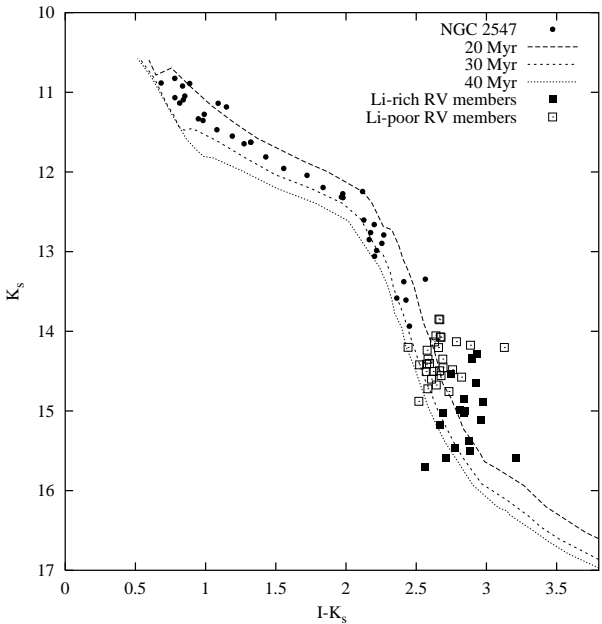}
\caption{A $K_{s}$ versus $I-K_{s}$ (2MASS system)  CMD for NGC 2547. Symbols and lines are as
  for Fig.~\ref{iritheory}
}
\label{kiktheory}
 \end{figure}

A purer test of the consistency between the LDB and low-mass isochronal
ages is available using the models of Baraffe et al. (2002). These
models, based on the evolutionary code of Chabrier \& Baraffe (1997),
predict both Li-depletion {\em and} the photospheric colours and
magnitudes using complex model atmospheres. 

Figure~\ref{ikldb} includes theoretical isochrones adjusted for an
intrinsic distance modulus of 8.1 mag and reddening of $E(I-K_{s})=0.092$.
These suggest an age of $25\pm 5$\,Myr for NGC 2547, but there are
additional errors due to the systematic photometric uncertainty of $\pm
0.1$ in the $I$ magnitudes of these cool stars, $\pm 0.031$ in $E(I-K_{s})$
and $\pm 0.1$ in the distance modulus. Hence the age from this method
is $25\pm 7$\,Myr, again in reasonable agreement with the LDB age found
from the same stars and evolutionary models. The $K_{s}$ versus $I-K_{s}$
diagram for stars in IC 2391 around its LDB suggests an age of
$40\pm10$\,Myr, also in reasonable agreement with the LDB age (see
Fig.~9b of Barrado y Navascu\'{e}s et al. 2004).

The cool stars we have considered here have rather imprecise 2MASS near
IR magnitudes. To improve on this we use the
spectroscopically confirmed sample of brighter NGC 2547 stars discussed
in section 6.3.2. The
comparisons between theory and data are shown in Figs.~\ref{iritheory},
\ref{iijtheory} and~\ref{kiktheory}. Note that the Baraffe et
al. (2002) colours and magnitudes on the CIT system have been converted
to the 2MASS system for these plots.

The agreement between the LDB age of NGC 2547 from the Chabrier \&
Baraffe (1997) models $(35\pm 3$\,Myr) and the theoretical
isochrone fits from the same models is also very good. The theoretical 
$I$ versus $R-I$ isochrones,
like the empirical isochrone modelling in Jeffries et al. (2004), suggest an age
of $30\pm5$\,Myr for $R-I<1.2$, the $I$ versus $I-J$ plot suggests an
age of $(30\pm 5)$\,Myr for $I-J<1.5$, whilst the $I$ versus $I-K_{s}$ CMD
suggests an age of $(25\pm 5)$ for $I-J<3$. In all the CMDs, but
particularly for $I$ versus $R-I$ there is a hint that the stars are
not arrayed parallel to the isochrones even in these colour ranges. At
redder colours this becomes more than a hint. The data and models are
clearly discrepant at any age for $R-I>1.2$ (a deficiency in the models already
noted by Baraffe et al. [1998] and probably due to missing atmospheric
molecular opacity in the optical spectrum of cool objects) and a drift
of stars towards younger ages for $I-J>1.5$ is also apparent. However,
recall that systematic errors of order 0.1 mag are possible in the
colours of the red stars so overall the theoretical NIR colours and
models do a reasonably good job of modelling the NGC 2547 data at or
just below the LDB age derived from the same model. It is interesting
to note though that the theoretical model atmospheres do not simulate
spots, plages, chromospheres or other manifestations of magnetic
activity in these young stars. Stauffer et al. (2003) found that cool
Pleiades stars seemed to show a small $K$-band excess with respect to
older stars. There is a hint of this also in our data, where the $K$
versus $I-K$ CMD gives a lower isochronal age than the other CMDs.
This would be brought back into line by the
addition of a $\simeq 0.1$ mag $I-K$ excess in the model atmospheres.

\section{Concluding Remarks}

Deep, intermediate resolution spectroscopy obtained with the
VLT/GIRAFFE spectrograph has been used to study a
group of 63 low-mass ($0.08-0.3\,M_{\odot}$) candidate members of the young open
cluster, NGC 2547. Radial velocity measurements and the presence of strong
\lii~6708\AA\ absorption 
in the fainter stars, have confirmed membership for at least
50 objects. We have hence validated previous work that
selected low-mass members and estimated the extent of sample
contamination based solely on photometric selection (Jeffries et
al. 2004). The conclusions of that work regarding the mass function and
luminosity function of NGC 2547 should be sound, at least down to the
substellar boundary.

From the sample of confirmed members we have determined the magnitude
at which lithium switches from being 
completely depleted, to being found
close to its initial abundance, as $I=17.54\pm0.14$, $K_{s}=14.86\pm0.12$
and at colours of $R-I=1.77\pm 0.03$ and $I-K_{s}=2.70\pm0.12$.

Using several PMS evolutionary models which incorporate a variety of
physical approximations we find an almost model-independent lithium
depletion boundary age (LDB) of 34-36\,Myr, with a precision of 10 per
cent. This LDB age is only slightly larger than the ages of 25-35 $(\pm
5)$\,Myr deduced from isochrone fits to stars from 0.08-1.2$M_{\odot}$
using the same models and empirically calibrated relationships between
colour, bolometric correction and $T_{\rm eff}$. It is also slightly
larger than ages of 25-30 ($\pm 5$)\,Myr deduced from isochrone fits in
the $I,I-J$ and $I,I-K_{s}$ CMDs using only the theoretical colours and
magnitudes of Baraffe et al. (2002).  However, perfect agreement with the LDB
ages can be obtained with a distance reduction to the cluster of only
0.1 mag, approximately equal to its estimated error bar.

On the basis that the isochrone fits to low-mass are compatible with
LDB ages derived from the same models we find that all of the PMS
models we have considered (Chabrier \& Baraffe 1997; D'Antona \&
Mazzitelli 1997; Siess et al. 2000, Baraffe et al. 2002) give
acceptable results. This is despite possible systematic uncertainties
regarding the intrinsic colours of cool, magnetically active young
stars (Stauffer et al. 2003).  At the age of NGC 2547, significant
differences in the isochrones predicted by these models occur at
colours so red that uncertainties in photometric calibration prevent
any decisive test between them. However, this result now seems to
remove any pressing need to introduce new physics such as rotation or
extra magnetic support into the PMS models and lends confidence to
isochronal ages determined (using the methods outlined here) 
from low-mass stars between 30 and 120\,Myr.
This is an important result, because PMS isochronal ages can
easily be determined from photometry obtained on small telescopes,
whereas even for nearby clusters, determining LDB ages stretches the capabilities of 
8-m telescopes.

The differences in the isochrones predicted by the various flavours of
model, arising from variations in their treatment of convection,
atmospheres, equations of state and interior opacities, become more
apparent at younger ages. This work demonstrates that finding the LDB
is certainly possible down to $I\simeq 18.5$ and therefore the
technique could and should be applied to more distant clusters in the
10-20\,Myr age range.

\section*{Acknowledgments}
We would like to thank the staff of the VLT in Paranal for delivering a
superb dataset in service mode. Data reduction was performed on
computing facilities at Keele University funded by the UK Particle
Physics and Astronomy Research Council (PPARC). JMO acknowledges
financial support from PPARC.  We thank Tim Naylor for useful
discussions. This publication makes use of data products from the Two
Micron All Sky Survey, which is a joint project of the University of
Massachusetts and the Infrared Processing and Analysis
Center/California Institute of Technology, funded by the National
Aeronautics and Space Administration and the National Science
Foundation.

\nocite{meynet93}
\nocite{kenyon04}
\nocite{houdebine95}
\nocite{kenyon04}
\nocite{stauffer03}
\nocite{young04}
\nocite{white03}
\nocite{cutri03}
\nocite{carpenter01}
\nocite{jeffries04}
\nocite{jeffries03}
\nocite{oliveira03}
\nocite{naylor02}
\nocite{jeffries98n2547}
\nocite{jeffries00}
\nocite{baraffe02}
\nocite{baraffe98}
\nocite{dantona97}
\nocite{leggett96}
\nocite{claria82}
\nocite{stauffer95}
\nocite{chabrier97}
\nocite{bessell79}
\nocite{kirkpatrick93}
\nocite{zapatero02}
\nocite{siess00}
\nocite{stauffer98}
\nocite{houdashelt00}
\nocite{pinsonneault98}
\nocite{barrado99}
\nocite{stauffer99}
\nocite{chiosi92}
\nocite{meynet97}
\nocite{meynet97}
\nocite{meynet00}
\nocite{bildsten97}
\nocite{ushomirsky98}
\nocite{jeffriescargese01}
\nocite{burke04}
\nocite{barradoldb04}
\nocite{blecha03}
\nocite{rieke85}
\nocite{dezeeuw99}
\nocite{pozzo00}
\nocite{pozzo01}
\nocite{gizis02}

\bibliographystyle{mn2e}  
\bibliography{iau_journals,master}

\begin{thebibliography}{}

\bibitem[\protect\citeauthoryear{Baraffe, Chabrier, Allard \&
  Hauschildt}{Baraffe et~al.}{1998}]{baraffe98}
Baraffe I.,  Chabrier G.,  Allard F.,    Hauschildt P.~H.,  1998, A\&A, 337,
  403

\bibitem[\protect\citeauthoryear{Baraffe, Chabrier, Allard \&
  Hauschildt}{Baraffe et~al.}{2002}]{baraffe02}
Baraffe I.,  Chabrier G.,  Allard F.,    Hauschildt P.~H.,  2002, A\&A, 382,
  563

\bibitem[\protect\citeauthoryear{Barrado~y Navascu\'es, Stauffer \&
  Jayawardhana}{Barrado~y Navascu\'es et~al.}{2004}]{barradoldb04}
Barrado~y Navascu\'es D.,  Stauffer J.~R.,    Jayawardhana R.,  2004, ApJ, in
  press

\bibitem[\protect\citeauthoryear{Barrado~y Navascu\'{e}s, Stauffer \&
  Patten}{Barrado~y Navascu\'{e}s et~al.}{1999}]{barrado99}
Barrado~y Navascu\'{e}s D.,  Stauffer J.~R.,    Patten B.~M.,  1999, ApJ, 522,
  L53

\bibitem[\protect\citeauthoryear{Bessell}{Bessell}{1979}]{bessell79}
Bessell M.,  1979, PASP, 91, 589

\bibitem[\protect\citeauthoryear{Bildsten, Brown, Matzner \&
  Ushomirsky}{Bildsten et~al.}{1997}]{bildsten97}
Bildsten L.,  Brown E.~F.,  Matzner C.~D.,    Ushomirsky G.,  1997, ApJ, 482,
  442

\bibitem[\protect\citeauthoryear{Blecha, North, Royer \& Simond}{Blecha
  et~al.}{2003}]{blecha03}
Blecha A.,  North P.,  Royer F.,    Simond G.,  2003, Technical report, GIRAFFE
  BLDR Software Reference Manual Version 1.09.
Observatoire de Gen\`eve

\bibitem[\protect\citeauthoryear{Burke, Pinsonneault \& Sills}{Burke
  et~al.}{2004}]{burke04}
Burke C.~J.,  Pinsonneault M.~H.,    Sills A.,  2004, ApJ, 604, 272

\bibitem[\protect\citeauthoryear{Carpenter}{Carpenter}{2001}]{carpenter01}
Carpenter J.~M.,  2001, AJ, 121, 2851

\bibitem[\protect\citeauthoryear{Chabrier \& Baraffe}{Chabrier \&
  Baraffe}{1997}]{chabrier97}
Chabrier G.,  Baraffe I.,  1997, A\&A, 327, 1039

\bibitem[\protect\citeauthoryear{Chiosi, Bertelli \& Bressan}{Chiosi
  et~al.}{1992}]{chiosi92}
Chiosi C.,  Bertelli G.,    Bressan A.,  1992, ARA\&A, 30, 235

\bibitem[\protect\citeauthoryear{Clari\'{a}}{Clari\'{a}}{1982}]{claria82}
Clari\'{a} J.~J.,  1982, A\&AS, 47, 323

\bibitem[\protect\citeauthoryear{{Cutri, R. M. et al.}}{{Cutri, R. M. et
  al.}}{2003}]{cutri03}
{Cutri, R. M. et al.} 2003, Technical report, Explanatory supplement to the
  2MASS All Sky data release.
http://www.ipac.caltech.edu/2mass/

\bibitem[\protect\citeauthoryear{D'Antona \& Mazzitelli}{D'Antona \&
  Mazzitelli}{1997}]{dantona97}
D'Antona F.,  Mazzitelli I.,  1997, Mem. Soc. Astr. It., 68, 807

\bibitem[\protect\citeauthoryear{de Zeeuw, Hoogerwerf, de Bruijne, Brown \&
  Blaauw}{de~Zeeuw et~al.}{1999}]{dezeeuw99}
de Zeeuw P.~T.,  Hoogerwerf R.,  de Bruijne J. H.~J.,  Brown A. G.~A.,
  Blaauw A.,  1999, AJ, 117, 354

\bibitem[\protect\citeauthoryear{Gizis, Reid \& Hawley}{Gizis
  et~al.}{2002}]{gizis02}
Gizis J.~E.,  Reid I.~N.,    Hawley S.~L.,  2002, AJ, 123, 3356

\bibitem[\protect\citeauthoryear{Houdashelt, Bell \& Sweigart}{Houdashelt
  et~al.}{2000}]{houdashelt00}
Houdashelt M.~L.,  Bell R.~A.,    Sweigart A.~V.,  2000, AJ, 119, 1448

\bibitem[\protect\citeauthoryear{Houdebine \& Doyle}{Houdebine \&
  Doyle}{1995}]{houdebine95}
Houdebine E.~R.,  Doyle J.~G.,  1995, A\&A, 302, 861

\bibitem[\protect\citeauthoryear{Jeffries \& Naylor}{Jeffries \&
  Naylor}{2001}]{jeffriescargese01}
Jeffries R.~D.,  Naylor T.,  2001, in Montmerle T.,  Andr\'{e} P.,  eds, From
  darkness to light: Origin and evolution of young stellar clusters ASP
  Conference Series, Vol. 243, San Francisco, p.~633

\bibitem[\protect\citeauthoryear{Jeffries, Naylor, Devey \& Totten}{Jeffries
  et~al.}{2004}]{jeffries04}
Jeffries R.~D.,  Naylor T.,  Devey C.~R.,    Totten E.~J.,  2004, MNRAS, 351,
  1401

\bibitem[\protect\citeauthoryear{Jeffries, Oliveira, Barrado~y Navascu\'{e}s \&
  Stauffer}{Jeffries et~al.}{2003}]{jeffries03}
Jeffries R.~D.,  Oliveira J.~M.,  Barrado~y Navascu\'{e}s D.,    Stauffer
  J.~R.,  2003, MNRAS, 343, 1271

\bibitem[\protect\citeauthoryear{Jeffries \& Tolley}{Jeffries \&
  Tolley}{1998}]{jeffries98n2547}
Jeffries R.~D.,  Tolley A.~J.,  1998, MNRAS, 300, 331

\bibitem[\protect\citeauthoryear{Jeffries, Totten \& James}{Jeffries
  et~al.}{2000}]{jeffries00}
Jeffries R.~D.,  Totten E.~J.,    James D.~J.,  2000, MNRAS, 316, 950

\bibitem[\protect\citeauthoryear{Kenyon, Jeffries, Naylor, Oliveira \&
  Maxted}{Kenyon et~al.}{2004}]{kenyon04}
Kenyon M.~K.,  Jeffries R.~D.,  Naylor T.,  Oliveira J.~M.,    Maxted P. F.~L.,
   2004, MNRAS, in press

\bibitem[\protect\citeauthoryear{Kirkpatrick, Kelly, Rieke, Liebert, Allard \&
  Wehrse}{Kirkpatrick et~al.}{1993}]{kirkpatrick93}
Kirkpatrick D.,  Kelly D.,  Rieke G.,  Liebert J.,  Allard F.,    Wehrse R.,
  1993, ApJ, 402, 643

\bibitem[\protect\citeauthoryear{Leggett, Allard, Berriman, Dahn \&
  Hauschildt}{Leggett et~al.}{1996}]{leggett96}
Leggett S.~K.,  Allard F.,  Berriman G.,  Dahn C.~C.,    Hauschildt P.~H.,
  1996, ApJS, 104, 117

\bibitem[\protect\citeauthoryear{Meynet \& Maeder}{Meynet \&
  Maeder}{1997}]{meynet97}
Meynet G.,  Maeder A.,  1997, A\&A, 321, 465

\bibitem[\protect\citeauthoryear{Meynet \& Maeder}{Meynet \&
  Maeder}{2000}]{meynet00}
Meynet G.,  Maeder A.,  2000, A\&A, 361, 101

\bibitem[\protect\citeauthoryear{Meynet, Mermilliod \& Maeder}{Meynet
  et~al.}{1993}]{meynet93}
Meynet G.,  Mermilliod J.~C.,    Maeder A.,  1993, A\&AS, 98, 477

\bibitem[\protect\citeauthoryear{Naylor, Totten, Jeffries, Pozzo, Devey \&
  Thompson}{Naylor et~al.}{2002}]{naylor02}
Naylor T.,  Totten E.~J.,  Jeffries R.~D.,  Pozzo M.,  Devey C.~R.,    Thompson
  S.~A.,  2002, MNRAS, 335, 291

\bibitem[\protect\citeauthoryear{Oliveira, Jeffries, Devey, Barrado~y
  Navascu\'{e}s, Naylor, Stauffer \& Totten}{Oliveira
  et~al.}{2003}]{oliveira03}
Oliveira J.~M.,  Jeffries R.~D.,  Devey C.~R.,  Barrado~y Navascu\'{e}s D.,
  Naylor T.,  Stauffer J.~R.,    Totten E.~J.,  2003, MNRAS, 342, 651

\bibitem[\protect\citeauthoryear{Pinsonneault, Stauffer, Soderblom, King \&
  Hanson}{Pinsonneault et~al.}{1998}]{pinsonneault98}
Pinsonneault M.~H.,  Stauffer J.~R.,  Soderblom D.~R.,  King J.~R.,    Hanson
  R.~B.,  1998, ApJ, 504, 170

\bibitem[\protect\citeauthoryear{Pozzo, Jeffries, Naylor, Totten, Harmer \&
  Kenyon}{Pozzo et~al.}{2000}]{pozzo00}
Pozzo M.,  Jeffries R.~D.,  Naylor T.,  Totten E.~J.,  Harmer S.,    Kenyon M.,
   2000, MNRAS, 313, L23

\bibitem[\protect\citeauthoryear{Pozzo, Jeffries, Naylor, Totten, Harmer,
  Kenyon \& Walter}{Pozzo et~al.}{2001}]{pozzo01}
Pozzo M.,  Jeffries R.~D.,  Naylor T.,  Totten E.~J.,  Harmer S.,  Kenyon M.,
   Walter F.~M.,  2001, in Montmerle T.,  Andr\'{e} P.,  eds, From darkness to
  light: Origin and evolution of young stellar clusters ASP Conference Series,
  Vol. 243, San Francisco, p.~801

\bibitem[\protect\citeauthoryear{Rieke \& Lebofsky}{Rieke \&
  Lebofsky}{1985}]{rieke85}
Rieke G.~H.,  Lebofsky M.~J.,  1985, ApJ, 288, 618

\bibitem[\protect\citeauthoryear{Siess, Dufour \& Forestini}{Siess
  et~al.}{2000}]{siess00}
Siess L.,  Dufour E.,    Forestini M.,  2000, A\&A, 358, 593

\bibitem[\protect\citeauthoryear{Stauffer, Barrado~y Navascu\'{e}s, Bouvier,
  Morrison, Harding, Luhman, Stanke, McCaughrean, Terndrup, Allen \&
  Assouad}{Stauffer et~al.}{1999}]{stauffer99}
Stauffer J.~R.,  Barrado~y Navascu\'{e}s D.,  Bouvier J.,  Morrison H.~L.,
  Harding P.,  Luhman K.,  Stanke T.,  McCaughrean M.,  Terndrup D.~M.,  Allen
  L.,    Assouad P.,  1999, ApJ, 527, 219

\bibitem[\protect\citeauthoryear{Stauffer, Hartmann \& Barrado~y
  Navascu\'es}{Stauffer et~al.}{1995}]{stauffer95}
Stauffer J.~R.,  Hartmann L.~W.,    Barrado~y Navascu\'es D.,  1995, ApJ, 454,
  910

\bibitem[\protect\citeauthoryear{Stauffer, Jones, Backman, Hartmann, Barrado~y
  Navascu\'{e}s, Pinsonneault, Terndrup \& Muench}{Stauffer
  et~al.}{2003}]{stauffer03}
Stauffer J.~R.,  Jones B.~F.,  Backman D.,  Hartmann L.~W.,  Barrado~y
  Navascu\'{e}s D.,  Pinsonneault M.~H.,  Terndrup D.~M.,    Muench A.~A.,
  2003, AJ, 126, 833

\bibitem[\protect\citeauthoryear{Stauffer, Schultz \& Kirkpatrick}{Stauffer
  et~al.}{1998}]{stauffer98}
Stauffer J.~R.,  Schultz G.,    Kirkpatrick J.~D.,  1998, ApJ, 499, L199

\bibitem[\protect\citeauthoryear{Ushomirsky, Matzner, Brown, Bildsten, Hilliard
  \& Schroeder}{Ushomirsky et~al.}{1998}]{ushomirsky98}
Ushomirsky G.,  Matzner C.~D.,  Brown E.~F.,  Bildsten L.,  Hilliard V.~G.,
  Schroeder P.~C.,  1998, ApJ, 497, 253

\bibitem[\protect\citeauthoryear{White \& Basri}{White \&
  Basri}{2003}]{white03}
White R.~J.,  Basri G.,  2003, ApJ, 582, 1109

\bibitem[\protect\citeauthoryear{Young E. T. amd~Lada, Teixeira, Muzerolle,
  Muench, Stauffer, Beichman, Rieke, Hines, Su, Engelbracht, Gordon, Misselt,
  Morrison, Stansberry \& Kelly}{Young et~al.}{2004}]{young04}
Young E. T. amd~Lada C.~J.,  Teixeira P.,  Muzerolle J.,  Muench A.,  Stauffer
  J.,  Beichman C.~A.,  Rieke G.~H.,  Hines D.~C.,  Su K. Y.~L.,  Engelbracht
  C.~W.,  Gordon K.~D.,  Misselt K.,  Morrison J.,  Stansberry J.,    Kelly D.,
   2004, ApJS, 154, 428

\bibitem[\protect\citeauthoryear{Zapatero~Osorio, B\'{e}jar, Pavlenko, Rebolo,
  Allende~Prieto, Mart\'{\i}n \& Garc\'{i}a~L\'{o}pez}{Zapatero~Osorio
  et~al.}{2002}]{zapatero02}
Zapatero~Osorio M.~R.,  B\'{e}jar V. J.~S.,  Pavlenko Y.,  Rebolo R.,
  Allende~Prieto C.,  Mart\'{\i}n E.~L.,    Garc\'{i}a~L\'{o}pez R.~J.,  2002,
  A\&A, 384, 937

\end{thebibliography}


\bsp 

\label{lastpage}

\end{document}